%% file: main.tex
\documentclass[letterpaper]{article} 
\usepackage{aaai25}  
\usepackage{times}  
\usepackage{helvet}  
\usepackage{courier}  
\usepackage[hyphens]{url}  
\usepackage{graphicx} 
\usepackage{natbib}  
\usepackage{caption} 
\usepackage{algorithm}
\usepackage{algorithmic}

\usepackage{newfloat}
\usepackage{listings}
\DeclareCaptionStyle{ruled}{labelfont=normalfont,labelsep=colon,strut=off} 
\floatstyle{ruled}
\newfloat{listing}{tb}{lst}{}
\floatname{listing}{Listing}
\title{Join the Chat: How Curiosity Sparks Participation in Telegram Groups}
\author {
    Giordano Paoletti\textsuperscript{\rm 1},
    Jussara M. Almeida\textsuperscript{\rm 2},
    Luca Vassio\textsuperscript{\rm 1},
    Marcos André Gonçalves\textsuperscript{\rm 2},
    Marco Mellia\textsuperscript{\rm 1}
    }
\affiliations {
    \textsuperscript{\rm 1} Politecnico di Torino, Turin, Italy\\
    \textsuperscript{\rm 2} Universidade Federal de Minas Gerais, Belo Horizonte, Brazil\\
    \{name.surname\}@polito.it, \{jussara,mgoncalv\}@dcc.ufmg.br
}

\usepackage[dvipsnames]{xcolor}
\usepackage{amsmath}
\usepackage{comment}
\definecolor{lincolngreen}{rgb}{0.11, 0.35, 0.02}

\newcommand{\answerYes}[1]{\textcolor{blue}{#1}} 
\newcommand{\answerNo}[1]{\textcolor{teal}{#1}} 
\newcommand{\answerNA}[1]{\textcolor{gray}{#1}}

\usepackage[utf8]{inputenc} 
\usepackage[T1]{fontenc}    
\usepackage{url}            
\usepackage{booktabs}       
\usepackage{amsfonts}       
\usepackage{nicefrac}       
\usepackage{microtype}      
\usepackage{lipsum}
\usepackage{fancyhdr}       
\usepackage{graphicx}       
\graphicspath{{media/}}     
\usepackage{multirow}
\usepackage{subcaption}

\begin{document}
\maketitle

\begin{abstract}
This study delves into the mechanisms that spark user curiosity driving \textit{active} engagement within public Telegram groups. By analyzing approximately 6 million messages from 29,196 users across 409 groups, we identify and quantify the key factors that stimulate users to actively participate (i.e., send messages) in group discussions. These factors include social influence, novelty, complexity, uncertainty, and conflict, all measured through metrics derived from message sequences and user participation over time. 
After clustering the messages, we apply explainability techniques to assign meaningful labels to the clusters. This approach uncovers macro categories representing distinct curiosity stimulation profiles, each characterized by a unique combination of various stimuli.
Social influence from peers and influencers drives engagement for some users, while for others, rare media types or a diverse range of senders and media sparks curiosity.
Analyzing patterns, we found that user curiosity stimuli are mostly stable, but, as the time between the initial message increases, curiosity occasionally shifts.
A graph-based analysis of influence networks reveals that users motivated by direct social influence tend to occupy more peripheral positions, while those who are not stimulated by any specific factors are often more central, potentially acting as initiators and conversation catalysts.
These findings contribute to understanding information dissemination and spread processes on social media networks, potentially contributing to more effective communication strategies.\looseness=-1
\end{abstract}


\section{Introduction}

Telegram is an instant messaging platform that has gained significant popularity since its launch in 2013. As of July 2024, Telegram has more than 950 million monthly active users\footnote{\textit{``Du Rove's Channel''}. \url{https://t.me/durov/337}, 22 July 2024. Accessed \today.}.
One of Telegram distinctive features is its support for large groups 
that can accommodate up to 200,000 members 
making it a powerful tool for discussion within a vast audience.\looseness=-1 




In this complex scenario, studying the dynamics that drive user engagement - i.e., the active posting of content in groups - is key to understanding the processes that guide information dissemination.  
Prior studies have analyzed {\it user's engagement} in social media platforms \cite{aldous2019view,vassio2022mining} using various metrics (e.g., number of likes, comments, shares) to capture how users {\it consume content and interact with others}. Specifically on Telegram,  \citet{Hashemi_Zare_Chahooki_2019} proposed indicators to distinguish high-quality from low-quality groups, marked by irregular activity, frequent spam, and unclear topics. \citet{hoseini2024characterizing} studied the dynamics of information dissemination in groups, examining message forwarding and lifespan.\looseness=-1

The large range of features makes Telegram an interesting platform for studying online social movements, extremism, and misinformation  spread~\cite{baumgartner_pushshift_2020, Urman_Katz_2022, venancioICWSM2024,kloo2024cross}. 
Evidence suggests that some groups even use Telegram to coordinate illicit activities, such as pump-and-dump schemes in the cryptocurrency market \cite{xu_anatomy_2019}, as well as form pods where members interact with each other’s content to artificially boost the popularity of their Instagram accounts \cite{weerasinghe_pod_2020}.\looseness=-1

We here look into user engagement on Telegram from a novel and complementary point of view: we study 
what drives user participation in group discussions by posting content,  rather than as a passive observer. We call this \textit{curiosity}. It is a fundamental human cognition element that, when adequately stimulated, becomes a driving force behind a person's choice to consume and, in the present case, {\it to produce information}~\cite{Sousa_Almeida_Figueiredo_2022b}. As such, by focusing on factors that stimulate user curiosity to join discussions, we offer a novel perspective of the processes contributing to information spread on online media platforms such as Telegram is.\looseness=-1

Human curiosity modeling finds its roots in the Psychology domain, with theories such as the  Information Gap Theory \cite{loewenstein1994psychology, kidd2015psychology}, which tries to explain how one's curiosity responds to external stimuli, as one of its foundational pillars. 
Also, according to the literature \cite{berlyne1960conflict},  one's curiosity stimulation process has many facets, manifested as stimulus components, referred to as {\it collative variables}. In Computer Science, a number of recent studies have proposed metrics to operationalize (some of) those variables in various scenarios,  often using them as components of information services, notably recommendation systems (e.g., \citet{fu2023modeling,xu2019enhancing}).\looseness=-1

In this work, we aim at studying {\it curiosity stimulation driving user participating in discussions} in public Telegram groups. 
Our investigation builds upon the work of \citet{Sousa_Almeida_Figueiredo_2022b}, which proposed metrics to measure curiosity stimulation in Brazilian politically oriented WhatsApp groups. 
Our study extends and diverges significantly from this prior work in four key aspects: \\
(i) The prior work explored only metrics related to ``social influence''. Instead, here we take a comprehensive approach by considering multiple complementary facets of curiosity stimuli, which allows us to fully characterize user's curiosity and its role in driving their participation in group discussion.\\
%
(ii) Unlike prior work, which was limited to a single topic (politics) within a specific cultural and linguistic context (Brazilian election), our study spans diverse topics, 
across multiple groups, and a heterogeneous population. This broad scope mitigates biases from local effects, enhancing the generalization of our findings and enabling the exploration of how curiosity profiles vary across different discussion topics.\\
(iii) We analyze the connection between curiosity and the user's engagement (i.e., the number of messages they post), a critical factor for understanding the role of curiosity in information spread that was not considered in the prior work. This analysis sheds light on how curiosity influences user activity participation in group discussions.\\
(iv) Telegram groups offer a unique and novel environment for this analysis due to their much larger membership\footnote{Telegram allows at most 200 thousand users in any group, whereas the Whatsapp groups were limited to 250 users in the prior work (the limit has since then grown to 1000 users).}, 
which creates a richer and more varied set of potential sources of curiosity stimulation driving the discussions.\looseness=-1



Our study is guided by the following research questions:
\begin{enumerate}
    \item \textbf{RQ1:} Which curiosity \textit{stimuli} 
    more often drive user participation? Do the stimuli experienced by users change over time as they join the group discussions?
    Does the discussion topic affect the prevalence of certain stimuli?\looseness=-1
    \item \textbf{RQ2} 
    What types of \textit{stimuli} activate the most engaged users? What types relate to those who are most influential in initiating and perpetuating conversations?
\end{enumerate}




Regarding RQ1, our analysis identifies six major \textit{curiosity stimulation profiles}.
Users generally maintain stable curiosity profiles, though, over time, they may shift between profiles in response to varying stimuli. 
The distribution of curiosity stimulation profiles varies across topics. For instance, users in political groups are mostly motivated by diverse and balanced discussions. To answer RQ2, the impact of specific profiles on participation and influence is notable: users drawn to direct peer interactions often engage less, while those motivated by uncertainty are more active. Users without specific curiosity-driven stimuli tend to hold central roles in discussions, often initiating and steering conversations.\looseness=-1


In the following, we introduce the background on curiosity stimulation and the metrics to capture its main components (Sec. \textit{Curiosity Stimulation}). We then describe the data (Sec. \textit{Telegram Groups Dataset}) we use in our study and discuss the main results (Sec. \textit{Analysis of Curiosity Stimulation}). Finally, we offer conclusions and directions for future work (Sec. \textit{Conclusions and Future Work}). Appendix A delivers additional results.\looseness=-1

\section{Curiosity Stimulation}
\label{sec:CuriosityMetrics}

\subsection{Background}

Previous works on Psychology have provided theoretical background to better understand human curiosity.
According to \cite{loewenstein1994psychology}, our organism receives various stimuli from the environment and selectively responds to those that generate more pleasure and raise our curiosity. \citet{berlyne1960conflict} identified four collative variables pertaining to the process of curiosity stimulation: {\it novelty}, {\it complexity}, {\it uncertainty}, and {\it conflict}.
These variables have been extensively validated in Psychology as main drivers of human curiosity and engagement, forming a cornerstone of modern cognitive and behavioral research field. The author argued that these variables can be quantified using information theory metrics, an idea later reinforced by \citet{silvia2006exploring}, who demonstrated the mathematical feasibility of such quantifications. This formalization bridges theoretical constructs and quantifiable metrics, ensuring a rigorous grounding in Psychology while enabling computational applications.\looseness=-1

Inspired by these concepts, recent Computer Science studies proposed metrics to capture one or more of these variables, using them as components of information services (e.g., recommendation functions) \cite{zhao2016much,xu2019enhancing, fu2023modeling, yan2023influence,
deng2023curiosity, wang2023item, tang2022preference}.
\citet{sousa2019analyzing} presented one of the first efforts to operationalize all four of Berlyne's collative variables to study curiosity stimulation in online music consumption. 
Similarly, \citet{fu2023modeling} proposed metrics based on information theoretical measures,
following the principles argued by \citet{berlyne1960conflict} and \citet{silvia2006exploring}.  By leveraging these well-established principles, these prior studies demonstrated the validity and reliability 
of Information Theory metrics in operationalizing curiosity stimulation across diverse contexts.
These metrics are quantified in bits, facilitating direct comparisons across systems \cite{timme2018tutorial}.  

Subsequently, \citet{Sousa_Almeida_Figueiredo_2022b} introduced metrics to quantify curiosity stimulation that motivates users to share content in public, politically oriented WhatsApp groups. They emphasized the importance of incorporating \textit{social influence} as a fifth factor in curiosity stimulation, complementing the four collative variables previously discussed. 
The inclusion of social influence aligns with the broader psychological framework defined by \citet{KASHDAN2018130}, which emphasizes the exploratory behaviors and interpersonal dynamics driving curiosity in group settings. Following the findings of \citet{ver2013information}, which demonstrated the utility of information-theoretical metrics for measuring social influence, they proposed metrics to account for all these factors. 
Nevertheless, their analysis focused exclusively on social influence metrics to identify \textit{profiles} of curiosity stimulation.
Our work adopts these metrics to explore curiosity stimulation within the domain of Telegram group discussions, building on the psychological and computational insights from these prior studies.\looseness=-1

Our goal is to study curiosity stimulation as a key factor which drives {\it users to actively engage in Telegram group discussions} by posting messages i.e., from the perspective of {\it content production}.  This contrasts with a body of studies of user engagement from the perspective of content consumption  (e.g., \citet{shivaram2024forecasting,  aldous2019view, vassio2021temporal}).
Our contribution lies in conducting a comprehensive analysis of all factors influencing curiosity stimulation, investigating a broader range of topics across a heterogeneous user population, and exploring how curiosity stimulation operates in the larger user base of Telegram groups, which differs from the smaller user bases in previous studies. In addition, for the first time, we study the relationship between curiosity and user engagement.\looseness=-1  

\subsection{Curiosity Stimulation Metrics}
\label{subsec:Metrics of Curiosity}

One's curiosity is constantly stimulated by signals from the environment \cite{loewenstein1994psychology}. In this work, we study the user curiosity stimulation within a given Telegram group. 
Our key assumption is that the messages recently posted in the group act as stimulus for driving a user to post a message, i.e., past messages stimulate the user's curiosity.\footnote{If the same user is active in multiple groups, their curiosity stimulation is analyzed separately and independently for each group.}
As such, we model each Telegram group as the sequence of messages posted over time and assume that stimulation may evolve as users post messages. Thus, for a given group, we measure curiosity stimulation by computing metrics associated with the collative variables at each {\it message} (also referred to as \textit{posting event}, hereafter).
For this, we define the {\it window of interaction} as a time interval  $\Delta T$ preceding the current posting event. Based on all messages posted during $[t - \Delta T, t]$ in the group, we compute metrics of curiosity stimuli driving the user to post a message at time $t$. 
\looseness=-1


To select $\Delta T$, we build on prior studies of collective attention in WhatsApp groups which adopted a window of attention of 30 minutes \cite{caetano2021analyzing}. In the absence of a corresponding analysis for Telegram, we use explicit replies\footnote{On Telegram, a user can explicitly reply to a message by mentioning it.} as a proxy for attention, observing that 93.7\% of users have a median reply time below 30 minutes.  
The choice $\Delta T=30$ minutes ensures that 85.5\% of messages have at least 10 preceding messages within the same time window, meeting the minimum volume requirement for reliable metric computation (see next Section). Thus, setting $\Delta T$ to 30 minutes strikes 
a good balance between filtering noise from the user’s attention period and maintaining sufficient data coverage (see Fig. \ref{fig:reply_median_deltaT} in Appendix A).\looseness=-1

Towards computing the metrics, we represent each message by the (anonymized) sender ID, the timestamp, and the set of media types  (i.e., text, URL, image, video, audio, poll, or other\footnote{\textit{Other} aggregates all other media that can be posted on Telegram but were found to be rarely used (stickers, gifs, documents, etc.).}) in the message.  We refer to the latter as the media {\it categories} associated with the message. Note that we do not exploit the message content itself 
which may depend on the context (e.g., topic, language, slang) associated with a particular group. 
By only using macroscopic category properties, we aim to compare curiosity metrics across groups, language, topics, etc., irrespective of such contextual properties. Note that we can apply this approach even when the actual content is not available (e.g., due to privacy issues, or because the message itself contains no text).\looseness=-1 

To quantify the collative variables of (i) novelty, (ii) uncertainty, (iii) conflict, (iv) complexity and (v) social influence, 
we compute 9 metrics in total, extending the derivations proposed in \citet{Sousa_Almeida_Figueiredo_2022b}. 
Given a message posting event, we compute each metric 
taking either all the users who posted messages in $\Delta T$ (\textit{user} metrics) or the categories of media used in those messages (\textit{category} metrics). 
Table \ref{tab:metrics} provides an overview of the 9 metrics. 
Below, we summarize their operational formulation and intuition.\looseness=-1

\input{text/curiosity_metrics_new}

\section{Telegram Groups Dataset}
\label{sec:DataCollection}

\begin{figure}[tbh!]
    \centering
\begin{subfigure}[b]{0.85\linewidth}
        \centering
\includegraphics[width=\textwidth]{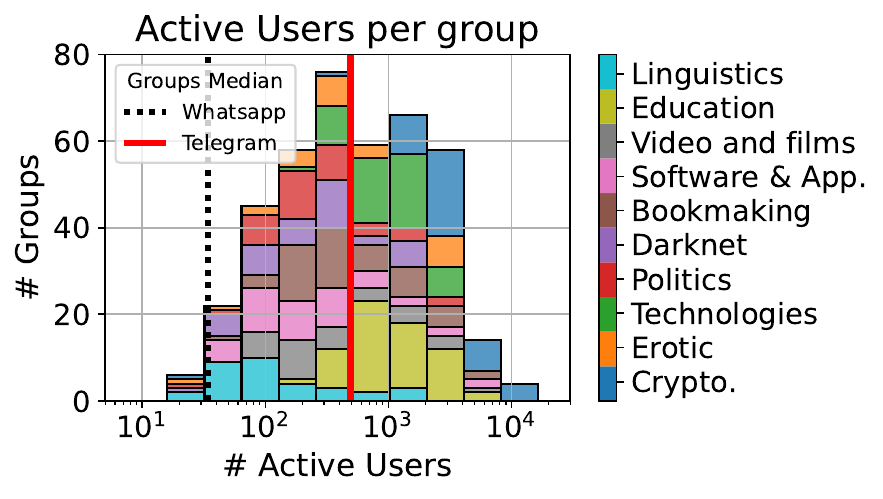}
        \caption{Distribution of the number of users (900k) who sent at least one message in the period under consideration among the 409 analysed groups .}
        \label{fig:active_users}

\end{subfigure}
    \begin{subfigure}[b]{0.75\linewidth}
        \centering
        \includegraphics[width=\textwidth]{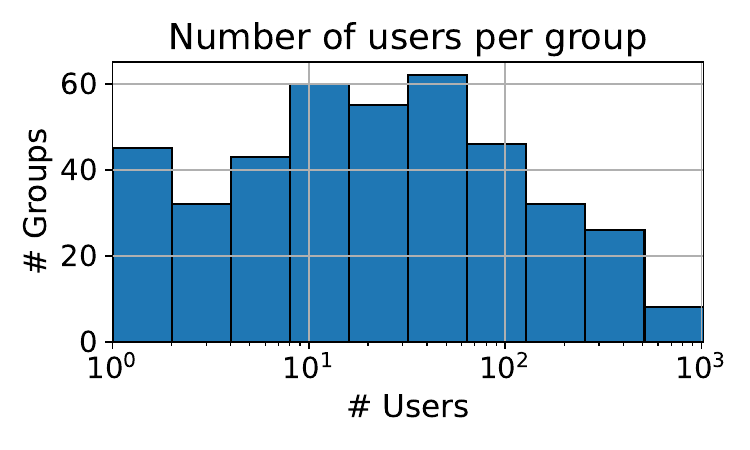}
        \caption{Distribution of the number of users per group for which we compute stimulus metrics (29k users).}
        \label{subfig:user_per_group}
    \end{subfigure}
   
    \begin{subfigure}[b]{0.75\linewidth}
        \centering
        \includegraphics[width=\textwidth]{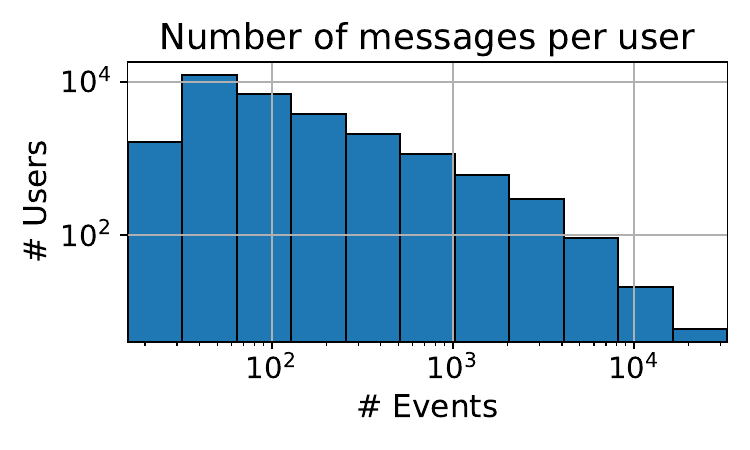}
        \caption{Distribution of the number of messages (6 million events) per user for which we compute stimulus metrics.}
        \label{subfig:event_per_user}
    \end{subfigure}
    

    \caption{Overview of analyzed Telegram dataset.}
    \label{fig:preliminar_distribution}
        \vspace{-0.3cm}
\end{figure}

Our study relies on a dataset collected by \citet{perlo2024topicwiseexplorationtelegramgroupverse}. The authors used \textit{TGStat}, an open catalogue of worldwide Telegram public groups\footnote{\url{https://tgstat.com}.}, to discover popular groups. Using the Telethon Python package \cite{telethon_api} the authors collected data from groups across 10  topics:  {\it Education, Bookmaking, Cryptocurrencies, Technologies, Darknet, Software and apps, Video and Films, Politics, Erotic} and {\it Linguistics}. 
Group topics were labelled by TGStat and manually validated by the authors. In Appendix A, Table \ref{tab:topic-description} offers a qualitative description of the topics.  We use a subset of the dataset that covers the whole month of March 2024, 
with around 22 million messages posted by more than 900 thousand users. \looseness=-1  
The authors kindly shared the data with us, which consists of the ordered sequence of messages for each group with anonymized user identifiers.

Although we cannot tell when each user became a member of the group, we can 
assess when they first appeared in our data. We found  that the average rate of first-time user 
appearances reaches 10\% on the first day before declining and stabilizing at roughly 2.5\% 
(see Figure \ref{fig:daily_activation_rate} in Appendix A).
Considering that our analysis focuses on active users (see below), we find that the average time between the first and last messages of these active users in our dataset is more than 15 days.
These observations suggest that, although the context surrounding the analyzed groups changes dynamically over time, the behavior of involved users does not change abruptly. Our focus is on users who engage consistently in group discussions over time, rather than those who exhibit a burst of activity and then disengage.\looseness=-1

As shown in Figure \ref{fig:active_users}, there is great diversity across the groups in terms of number of active users (i.e., users who posted at least one message). This diversity is also clear across the various topics.  For example, groups in \textit{Linguistics} tend to have fewer active users, whereas groups in \textit{Education} and \textit{Cryptocurrencies} tend to have larger numbers of users actively posting content. In particular, note the presence of many groups (35\%) with at least 1,000 active members, which is much larger than the limit of 250 simultaneous members (not necessarily actively posting content) allowed by WhatsApp when the dataset analyzed by \citet{Sousa_Almeida_Figueiredo_2022b} was gathered.   In fact, we note that the median number of active users in that dataset\footnote{Publicly available at \url{https://doi.org/10.5281/zenodo.5790153}}, computed 
during the month of peak activity (October 2018),  was only 34 users (as indicated by the dashed black vertical line in Fig. \ref{fig:active_users}), whereas here, groups have a median of 493 active users (red vertical line).
Such a large difference underscores the importance of studying curiosity on Telegram, given the novelty of its larger and more diverse social spaces.\looseness=-1  

To mitigate data sparsity issues, we follow \citet{Sousa_Almeida_Figueiredo_2022b} and compute the metrics only for users who posted at least 30 messages over the entire period and only for posting events with at least 10 messages within the \textit{window of interaction}.  Messages from service bots are considered in the computation of metrics, as they are inherently part of the conversation. Yet, the final analysis focused only on metrics computed for users.\looseness=-1 

After the filtering, we end up with roughly 6 million messages. 29,196 users were active in 409 distinct groups. 
As shown in Figs. \ref{subfig:user_per_group} and \ref{subfig:event_per_user}, there is great heterogeneity in terms of the number of users per group and the number of messages per user for which we can compute the metrics. Such diversity reflects the wide range of activity levels among users.  
However, the coverage across topics is fairly homogeneous as the number of groups per topic ranges from 25 for Erotic to 60 for Education.\looseness=-1

\input{text/04_Analysis_of_the_curiosity_stimuli}
\input{text/05_Conclusion}


\section{Limitations}
\begin{itemize}
    \item 
    This study relies on data from public Telegram groups identified by TGStat. While  we cannot guarantee that TGStat is bias-free, we note that it has been used in previous research\cite{tikhomirova2021community,Urman_Katz_2022,morgia_its_2023}, which supports its reliability as a data source. Our analysis is confined to the data collected from the selected Telegram groups, so we do not claim broad generalizability, such as to private or less active groups. Yet, the strength of our study lies in its large dataset, which includes over 6 million messages across 409 groups and spans 10 diverse topics—far exceeding the scope of prior work.
 
    \item  Metrics used in our study operationalize cognitive concepts. Thus validation is extremely hard, as it might involve a multi-disciplinary approach. Nevertheless, metric derivation followed procedures of~\citet{Sousa_Almeida_Figueiredo_2022b}, which, in turn,   followed the seminal works by \citet{berlyne1960conflict, silvia2006exploring}, which proposed a methodology to quantify collative variables related to curiosity stimulation based on information theoretical metrics. Thus, despite the lack of direct validation, the used metrics are rooted in the foundational theories of curiosity stimulation.
 

     \item Our analysis of curiosity stimulation is focused on a subset of users and messages (i.e., users who shared at least 30 messages in the period, and messages with at least 10 other messages in the window of interaction).  Whereas this filtering decision was made for the sake of robust metric computation, we argue that 
     removed users can be considered too inactive to be studied with respect to curiosity stimulation. 
     Similarly, the curiosity driven by the posting of messages after periods of low or no activity might be less dependent on these (few or none) previous messages. 
    \item We deliberately choose not to analyze the message content (other than the type of media). We do recognize that content is likely to play a direct role in shaping user curiosity. Yet, our choice allows for more universal comparison of curiosity metrics across different groups and even social network platforms, free from the influence of specific contexts such as the discussion topic or language. This strategy also ensures broader applicability, particularly when content is inaccessible due to privacy concerns, making the method both flexible and ethically sound.
\end{itemize}
\section{ Ethical Statement}

The data in this paper was collected and posted by  \citet{perlo2024topicwiseexplorationtelegramgroupverse}. The authors state that it was gathered using the Telethon API, which in turn utilizes the official Telegram API. The dataset includes messages from public groups from which the collectors have not been removed and where the admins have not set a time limit for viewing such messages.

\section{Acknowledgments}
This work has been partially supported by the Spoke 1 ``FutureHPC \& BigData" of ICSC --- Centro Nazionale di Ricerca in High-Performance-Computing, Big Data and Quantum Computing, funded by European Union --- NextGenerationEU. It has also been partially supported by {\it Conselho Nacional de Desenvolvimento Científico e Tecnológico }(CNPq) and {\it Fundação de Amparo à Pesquisa do Estado de Minas Gerais } (FAPEMIG), both in Brazil.

\bibliography{references}

\section{Paper Checklist}

\begin{enumerate}

\item For most authors...
\begin{enumerate}
    \item  Would answering this research question advance science without violating social contracts, such as violating privacy norms, perpetuating unfair profiling, exacerbating the socio-economic divide, or implying disrespect to societies or cultures?
    \answerYes{Yes}
  \item Do your main claims in the abstract and introduction accurately reflect the paper's contributions and scope?
    \answerYes{Yes}
   \item Do you clarify how the proposed methodological approach is appropriate for the claims made? 
    \answerYes{Yes, see the Analysis of Curiosity Stimulation.}
   \item Do you clarify what are possible artifacts in the data used, given population-specific distributions?
    \answerNA{NA}
  \item Did you describe the limitations of your work?
    \answerYes{Yes, see the Limitations.}
  \item Did you discuss any potential negative societal impacts of your work?
    \answerNA{NA}
      \item Did you discuss any potential misuse of your work?
    \answerNA{NA}
    \item Did you describe steps taken to prevent or mitigate potential negative outcomes of the research, such as data and model documentation, data anonymization, responsible release, access control, and the reproducibility of findings?
    \answerNA{NA}
  \item Have you read the ethics review guidelines and ensured that your paper conforms to them?
    \answerYes{Yes}
\end{enumerate}

\item Additionally, if your study involves hypotheses testing...
\begin{enumerate}
  \item Did you clearly state the assumptions underlying all theoretical results?
    \answerYes{Yes}
  \item Have you provided justifications for all theoretical results?
    \answerYes{Yes}
  \item Did you discuss competing hypotheses or theories that might challenge or complement your theoretical results?
    \answerNA{NA}
  \item Have you considered alternative mechanisms or explanations that might account for the same outcomes observed in your study?
    \answerNA{NA}
  \item Did you address potential biases or limitations in your theoretical framework?
    \answerYes{Yes, see the Limitations}
  \item Have you related your theoretical results to the existing literature in social science?
    \answerYes{Yes, partially. Although curiosity has never been studied on Telegram before, we compared our results with those obtained in Whatsapp.}
  \item Did you discuss the implications of your theoretical results for policy, practice, or further research in the social science domain?
    \answerYes{Yes}
\end{enumerate}

\item Additionally, if you are including theoretical proofs...
\begin{enumerate}
  \item Did you state the full set of assumptions of all theoretical results?
    \answerNA{NA}
	\item Did you include complete proofs of all theoretical results?
        \answerNA{NA}
\end{enumerate}

\item Additionally, if you ran machine learning experiments...
\begin{enumerate}
  \item Did you include the code, data, and instructions needed to reproduce the main experimental results (either in the supplemental material or as a URL)?
    \answerYes{Yes}
  \item Did you specify all the training details (e.g., data splits, hyperparameters, how they were chosen)?
    \answerYes{Yes}
     \item Did you report error bars (e.g., with respect to the random seed after running experiments multiple times)?
    \answerNA{NA}
	\item Did you include the total amount of compute and the type of resources used (e.g., type of GPUs, internal cluster, or cloud provider)?
    \answerNo{No}
     \item Do you justify how the proposed evaluation is sufficient and appropriate to the claims made? 
    \answerNA{NA}
     \item Do you discuss what is ``the cost`` of misclassification and fault (in)tolerance?
    \answerNA{NA}
  
\end{enumerate}

\item Additionally, if you are using existing assets (e.g., code, data, models) or curating/releasing new assets, \textbf{without compromising anonymity}...
\begin{enumerate}
  \item If your work uses existing assets, did you cite the creators?
    \answerYes{Yes}
  \item Did you mention the license of the assets?
    \answerNA{NA}
  \item Did you include any new assets in the supplemental material or as a URL?
    \answerNo{No}
  \item Did you discuss whether and how consent was obtained from people whose data you're using/curating?
    \answerNo{No, the code used to compute the curiosity metrics is public and the dataset of Telegram messages was kindly shared with us by the authors who are fully aware of its use for the purposes of this paper only }
  \item Did you discuss whether the data you are using/curating contains personally identifiable information or offensive content?
    \answerYes{Yes, see Ethical Statement.}
\item If you are curating or releasing new datasets, did you discuss how you intend to make your datasets FAIR?
\answerNA{NA}
\item If you are curating or releasing new datasets, did you create a Datasheet for the Dataset?
\answerNA{NA} 
\end{enumerate}

\item Additionally, if you used crowdsourcing or conducted research with human subjects, \textbf{without compromising anonymity}...
\begin{enumerate}
  \item Did you include the full text of instructions given to participants and screenshots?
    \answerNA{NA} 
  \item Did you describe any potential participant risks, with mentions of Institutional Review Board (IRB) approvals?
    \answerNA{NA} 
  \item Did you include the estimated hourly wage paid to participants and the total amount spent on participant compensation?
    \answerNA{NA} 
   \item Did you discuss how data is stored, shared, and deidentified?
   \answerNA{NA} 
\end{enumerate}

\end{enumerate}

\appendix

\input{text/appendix}


\end{document}

%% file: text/curiosity_metrics_new.tex
\subsection{Operational Definitions} 
\label{subsec:Mathematical_Operationalization}

\begin{table*}[!t]
\centering
\begin{small}
\caption{Metrics of curiosity stimulation computed at each posting event of a message $m$ sent by $u$, considering a window of interaction $\Delta T$ preceding the message. Refer to section \textit{Operational Definitions} for details on the terms in the equations.}

\begin{tabular}{p{1.3cm} c p{7cm} l}
\toprule
\textbf{Collative Variable} & \textbf{Metric} & \textbf{Description} & \textbf{Definition}\\ 
\midrule                       
\multirow{2}{*}{Novelty} & \textit{userNovelty} & 
Novelty in user $u$ participation wrt. others who posted during $\Delta T$ & 
                                        $ \begin{cases}
                                        -\log_2 (\frac{n_{v|t,g}}{\sum_{v} n_{u|t,g}}) , & \text{if } n_{u|t,g}>0 \\
                                        -\log_2 (1/|U_{t,g}|),              & \text{otherwise}
                                        \end{cases} 
                                        $
                                        \\
                            & \textit{catNovelty} & 
                            Novelty in the category of $m$ wrt. message categories posted during $\Delta T$ 
                                        &$
                                        \begin{cases}
                                            -\log_2 (P_{t,g}(C_m)) , & \text{if } P_{t,g}(C_m)>0\\
                                            -\log_2 (1/|C_{t,g}|),              & \text{otherwise}
                                        \end{cases}
                                        $\\ \hline 
\multirow{2}{*}{Uncertainty} & \textit{userUncertainty} & 
                            Variety (entropy) in user participation  during $\Delta T$
                            & 
                            $ -\sum_{v\in U_{g}}  \frac{n_{v|t,g}}{\sum_{v'} n_{{v'}|t,g}} \log_2 (\frac{n_{v|t,g}}{\sum_{v'} n_{v'|t,g}})
                            $\\ 
                            & \textit{catUncertainty} & 
                            Variety (entropy) in media category usage in messages posted during $\Delta T$
                            & $-\sum_{c\in C_{t,g}}  \frac{n_{c|t,g}}{\sum_{c'}n_{c'|t,g}}  \log_2 (\frac{n_{c|t,g}}{\sum_{c'}n_{c'|t,g}})
                            $        \\ \hline
\multirow{2}{*}{Conflict} & \textit{userConflict} & Diversity of distinct users posting messages during $\Delta T$ & 
$- \log_2 (\frac{1}{|U_{t,g}|})$
\\ 
                            & \textit{catConflict} & Diversity of distinct media categories used during $\Delta T$ 
                            &
                            $ - \log_2 (\frac{1}{|C_{t,g}|}) $
                            \\ \hline 
                            
\multirow{1}{*}{Complexity} & \textit{catComplex} & Complexity wrt. media categories  used during $\Delta T$ & $-\log_2(\frac{|C_{t,g}|}{|M|})$\\   \hline
\multirow{2}{1 cm}{Social
                Influence} 
                            & \textit{maxDirInf} & Maximum direct influence to $u$ from  users who posted during $\Delta T$ &
                            $\max_{o \in O_{d,t,g}}PMI^+_{t,g}(d,o)$\\ 
                            & \textit{maxIndInf} & Maximum indirect influence from users who posted during $\Delta T$ &
                            $\max_{o' \in O_{d,t,g}}MI^+_{t,g}(D,o') $\\                             
\bottomrule
\end{tabular}
\label{tab:metrics}
\end{small}
\end{table*}


The curiosity driving the posting of a message by a given user $u$ at a given time $t$ depends on the past 
messages posted in the group (including messages posted by $u$ themself) 
in the interaction window $[t ; t-\Delta T ]$. 

We define the sets of users, groups, and types of media as $U,G,M$ respectively. We do not assume access to message content and use only the different categories of media used to compose the message to represent it. 
Each message under analysis is then a tuple $(u, C_m , g, t)$ indicating that it was posted by user $u\in U$ in group $g \in G$ at time $t$ with media belonging to set $C_m \subseteq M$. Let $U_g$ be the set of users in group $g$. 
Given these definitions, we define the different collative metrics as follows:\looseness=-1     

\noindent
\paragraph{Novelty:}
The $userNovelty$ metric captures the novelty of the posting experience for user $u$. It reflects that users who have posted frequently in recent times find the discussed topic less novel since they are already engaged in the debate, which in turn affects their curiosity. The metric is detailed in Tab. \ref{tab:metrics}, where $n_{v|t,g}$ is the number of messages sent in the window of interaction $[t ; t-\Delta T ]$ by any user $v$ (including $u$) in group $g$.  
If no previous message is sent by $u$, we set the novelty to its maximum value possible, which corresponds to the uniform distribution of users, i.e. $-\log_2 (1/|U_{t,g}|)$ where $U_{t,g}$ is the set of distinct users who posted content in group g during the current window of interaction.\looseness=-1 

Similarly, we define $catNovelty$, a novelty metric related to the message categories, based on the surprisal (i.e., Shannon information content) associated with that variable. It captures how surprising it is that the message posted at time $t$ contains the media categories in $C_m$.  As shown in Tab. \ref{tab:metrics}, it depends on
$P_{t,g}(C_m)$, the average probability of the media categories posted in the message. This measure is estimated as $$P_{t,g}(C_m)=\frac{1}{|C_m|} \sum_{c\in C_m}\frac{n_{c|t,g}}{\sum_{c'\in C_{t,g}}n_{c'|t,g}} $$ 
where $C_{t,g}$ is the set of media categories of all messages posted during the window $\Delta T$. Both novelty metrics vary between 0 and $+\infty$.\looseness=-1

\noindent
\paragraph{Uncertainty:} Both novelty-related metrics focus on the message posted at time $t$. The aggregation of either metric for all messages posted during the window of interaction, using entropy, captures the uncertainty driving user curiosity. The idea behind these metrics is that the curiosity of the user may be more/less stimulated by the greater/less variety in the users posting messages (and in the categories of these messages) during the window of interaction. Thus, we define metrics of uncertainty related to users and categories as detailed in Tab. \ref{tab:metrics}. 
Both uncertainty metrics vary between 0 and $1$.\looseness=-1


\noindent
\paragraph{Conflict:}
Conflict measures the diversity in the stimulus.
The basic idea is that the distinct elements (categories/users) that appear in the window of interaction represent the different responses stimulating the curiosity of the user, and the strength of each response is captured by the probability of occurrence of each element (category/user). Thus, we define metrics of conflict related to users and categories to quantify the difference in terms of number of distinct users or distinct categories present in the window of interaction, as detailed in Table \ref{tab:metrics}.
Both conflicts vary between 0 and $+\infty$: more users (categories) lead to greater conflict, while a single user (category) leads to no conflict (i.e., conflict equal to 0).\looseness=-1


\noindent
\paragraph{Complexity:}
An alternative form of capturing the diversity of message categories in the current window of interaction is by exploiting the unique occurrences of the media types. This metric, also based on surprisal, captures how many distinct categories of messages are posted during the current window of interaction, out of the total number of categories $|M|$. Following the definition in Table \ref{tab:metrics}, 
$catComplex$ varies between 0 (when all categories are present in the time window) to $\log_2(|M|)$ (when only a single category is present).\looseness=-1  
We note that complexity and conflict are functions of each other:  $catComplex(t,g)= \log_2(\frac{1}{|C_{t,g}|})+\log_2(|M|))=-catConflict(t,g) +\log_2(|M|))$.\looseness=-1   





\noindent
\paragraph{Social Influence}:
Social Influence encompasses two types of stimuli - direct and indirect. \textit{Direct influence} measures the impact from peers, while \textit{indirect influence} refers to the impact from influencers or activists, i.e., users who often post content and sustain discussions by encouraging broader group participation rather than influencing any single user. Thus, we use metrics for both types of stimuli.
To distinguish between the user whose curiosity we are analyzing, i.e., user $u$ who posted a message at time $t$ in group $g$ and the other users who may stimulate $u$’s curiosity through social influence, we refer to the former as the destination $d$ and the latter as origins $o$ of social influence.\looseness=-1  

\noindent
We define:\\
    (i) Probability of observing an origin $o$ on group $g$: 
    $P_{t,g}(O=o)$;\\
    (ii) Probability of observing a destination $d$ on group $g$: 
    $P_{t,g}(D=d)$;\\
    (iii) Probability of observing a destination $d$ given an origin $o$:
    $P_{t,g}(D=d|O=o)$;\\
    (iv) Joint probability of observing a destination $d$ and an origin $o$: 
   $P_{t,g}(D=d,O=o)$.\\
All these probabilities are conditioned on the group $g$ and timestamp $t$ of the current message and can be estimated given the message sequence history up to $t$.\looseness=-1 

The {\it direct } social influence metric, defined in Tab. \ref{tab:metrics}, is based on the concept of Pointwise Mutual Information ($PMI$), which is computed for a particular destination $d$ and origin $o$ as:
$$ PMI_{t,g}(d,o)= \log_2 (\frac{P_{t,g}(d|o)}{P_{t,g}(d)})\text{ if } P_{t,g}(d)>0 
$$
or $0$ otherwise.
The $PMI$ of $d$ and $o$ measures the reduction in the uncertainty of destination $d$ posting a message due to the knowledge that $o$ also posted a message within the window of interaction.
It means that, based on historical patterns, the behavior of user $d$ in the group is influenced by the recent behavior of user $o$. 
Note that $PMI$ is negative if $d$ and $o$ occur less frequently than would be expected under the assumption of independent behavior, then we clip negative values to 0, i.e. $PMI^+_{t,g}(d,o) = max (PMI_{t,g}(d,o),0)$

Since the current interaction window may have multiple origins stimulating curiosity destination $d$, we compute the maximum {\it direct} social influence for all origins $o$ on $d$ as detailed in Tab. \ref{tab:metrics}, 
where $O_{d,t,g}$ is the set of origins at time t on group g for d. 
In addition to direct influence, we argue that some users may have a natural ability to influence others.
Users who often post content and, in doing so, keep the discussion going by driving others to also post content. We consider that, even in the absence of prior experiences such influencers may still stimulate the curiosity of a user (e.g., a newcomer) in the group. We identify these {\it indirect influencers} by searching for origins $o$ that tend to have a strong influence towards any destination $d'\in D$. Given this rationale, we define a metric based on the mutual information of all the destinations conditioned on a particular origin, that is: $$MI_{t,g}(D,o)=\sum_{d'\in U_g} P_{t,g}(d',o)PMI_{t,g}(d',o)$$
\looseness=-1 

Again, we clip negative values of mutual information at 0 (i.e., $MI^+$) as a reflection of no social influence from origin $o$ on any destination. Smaller values of $MI^+$ suggest weaker and no clear influence of $o$ over users. 
We also aggregate the metric for all origins in the window $\Delta T$ by taking the maximum,  as detailed in Tab. \ref{tab:metrics}.\looseness=-1  
 




Both social influence metrics vary between 0 and $+\infty$.
Note that \citet{Sousa_Almeida_Figueiredo_2022b}  proposed two variants of each social influence metric, based on the maximum and the average across all origins. Yet, they found both variants to be strongly correlated with each other, keeping only the maximum in their study. As such, we also choose to use only the maximum.\looseness=-1

%% file: text/04_Analysis_of_the_curiosity_stimuli.tex
\section{Analysis of Curiosity Stimulation}
\label{sec:Analysis}

\subsection{Metric Selection}

\begin{figure}[tbp!]
    \centering
    \includegraphics[width=0.8\linewidth]{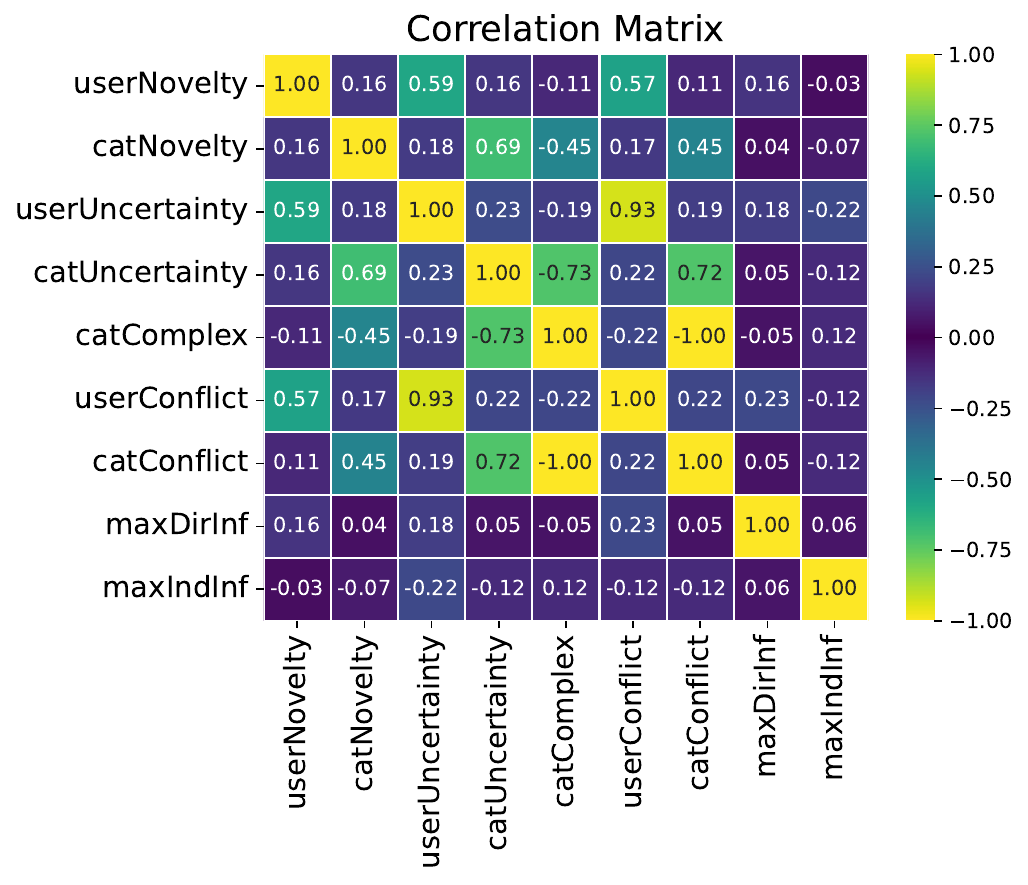}
    \caption{Correlation  among curiosity  metrics for all messages.
}
    \label{fig:fraction_corr}
        \vspace{-0.3cm}
\end{figure}

Recall that we have 9 metrics to capture components (collative variables) of curiosity stimulation. As initial step, we assess whether these metrics are correlated one to each other. 
To that end, we first compute all 9 metrics, standardize them according to a Z-score normalization, then compute the correlation matrix among the features (see Fig. \ref{fig:fraction_corr}). 
Our findings indicate that direct and indirect social influence metrics capture distinct aspects of social-related curiosity, and both metrics have weak correlations with the other metrics.
This shows that these metrics provide complementary dimensions of curiosity stimulation compared to the traditional collative variables. Among the other metrics, we find that: (i) {\it userConflict} and \textit{userUncertainty} are strongly correlated (Pearson's correlation $\rho$ above 0.8), and (ii) {\it catConflict} and \textit{catComplexity} has a perfect inverse correlation ($\rho=-1$), as expected from their formulation (see previous section). We then keep only one of these strongly correlated pairs of metrics. We end up with 7 complementary metrics, namely: \textit{userNovelty}, \textit{catNovelty}, \textit{userUncertainty}, \textit{catUncertainty},  \textit{catComplex},  \textit{maxIndInf} and  \textit{maxDirInf}.\looseness=-1

These observations align with previous findings for WhatsApp groups \cite{Sousa_Almeida_Figueiredo_2022b}. However, our study differs in a key methodological aspect: 
we consider all seven metrics to infer curiosity profiles in Telegram groups, while prior work used only two social influence metrics to infer curiosity profiles in WhatsApp groups. Our broader approach allows for a more comprehensive analysis, revealing richer and more diverse profiles, as discussed next.



\subsection{Message-Level Curiosity Stimulation}
\label{subsec:message}

\begin{table*}[htbp!]

\caption{Median of z-score normalized curiosity stimulus metrics for message clusters. An asterisk (*) indicates when a metric distribution is significantly higher than those in each other cluster (Mann-Whitney test, $p<0.01$). }
\label{tab:median_metrics}

\centering
\begin{small}
\begin{tabular}{r|c|rrrrrrr|r}
\toprule
ID& Description & \textit{userNov} & \textit{catNov} & \textit{userUncert} & \textit{catUncert} & \textit{catComplex} & \textit{maxDirInf} & \textit{maxIndInf}& \% Messages   \\
\midrule
0&Indirect Influence&-0.82&-0.56&-0.91&-0.92&1.30&-0.64&0.14*&14.8\%\\
1& Direct Influence &-0.23&-0.47&-0.06&-0.48&-0.08&1.47*&-0.08&13.2\%\\
2&Mixed stimuli &0.61&-0.40&0.43&-0.18&-0.08&-0.13&-0.34&19.8\%\\
3&Categorical Novelty&-0.15&2.93*&-0.29&1.28&-0.08&-0.16&-0.25&6.8\%\\
4&Independent&-0.83&-0.35&-0.58&-0.04&-0.08&-0.54&-0.25&21.7\%\\
5& Uncertainty&0.82*&0.98&1.21*&1.57*&-1.15&0.21&-0.45&11.3\%\\
6&Mixed stimuli &0.45&-0.56&0.55&-0.92&1.30*&-0.12&-0.36&12.5\%\\
\bottomrule
\end{tabular}
\end{small}
\vspace{-0.3cm}
\end{table*}

We start by investigating the curiosity stimuli that more often drive users to post a new message.
To that end, we infer profiles by clustering messages (independently from users and groups), according to the 7 identified metrics. 
Given the large dataset size (6 million messages), we employ an iterative stochastic clustering approach. We use MiniBatch K-means \cite{sculley2010web}, running 10 iterations for each tested number of clusters ($k$). We define the best number of clusters by identifying the elbow point in the average inertia curve, which measures the sum of squared distances of samples to their nearest cluster centroid. Following this method, we identify 7 clusters, each reflecting a different profile of curiosity stimulation.\looseness=-1  

We proceed to characterize the identified clusters with respect to any distinctive behavior in terms of metric values. To that end, we adopt a combined approach by comparing boxplots of metric distributions within each cluster and analyzing the SHAP (SHapley Additive exPlanations) values derived from a decision tree classifier \cite{lundberg2020local2global}. We omit the plots here due to space constraints (deferring them to Figs. \ref{fig:message_cluster_boxplots} and \ref{fig:main} in Appendix A). In turn,  Tab. \ref{tab:median_metrics}  offers a high-level description of each cluster through the median values of all metrics, highlighting those that are higher with statistical significance. We label each cluster ({\it Description} column of the table) according to those distinctive patterns.

Clusters 0 and 1 refer to messages whose postings were mostly stimulated by (indirect and direct) social influence. They account for roughly 28\% of all messages in our dataset. Clusters 3 and 5 consist of messages where the traditional collative variables, notably novelty and uncertainty, play a stronger role in curiosity stimulation. Clusters 2 and 6 have mixed stimuli. Given their similarities in terms of metric distributions  (and having the closest centroids), we choose to merge 
them into a single cluster (Cluster 2, mixed stimuli) to enhance interpretability. Finally, cluster 4 consists of messages for which no clear evidence of curiosity stimulation was found (i.e., curiosity independent).\looseness=-1  

Here we compare our findings with those of \citet{Sousa_Almeida_Figueiredo_2022b} that analyzed political groups in Whatsapp. For this, we restrict our analysis to Politics groups on Telegram. Results reveal striking differences: We find six significant clusters for Telegram while only three emerged in Whatsapp. On Telegram, Direct and indirect social influences dominate only 2.6\% and 8\% of the messages, versus 13\% and 14.4\% on WhatsApp. Messages with no clear curiosity stimulation make up only 21\% in our data, while those where 73\% in WhatsApp. These differences stem from the focus of the prior work restricted to social influence. By incorporating all curiosity metrics, we reveal other profiles that are also important for Politics groups.





\begin{figure}[ttt!]
    \centering
    \includegraphics[width=
   0.8\linewidth]{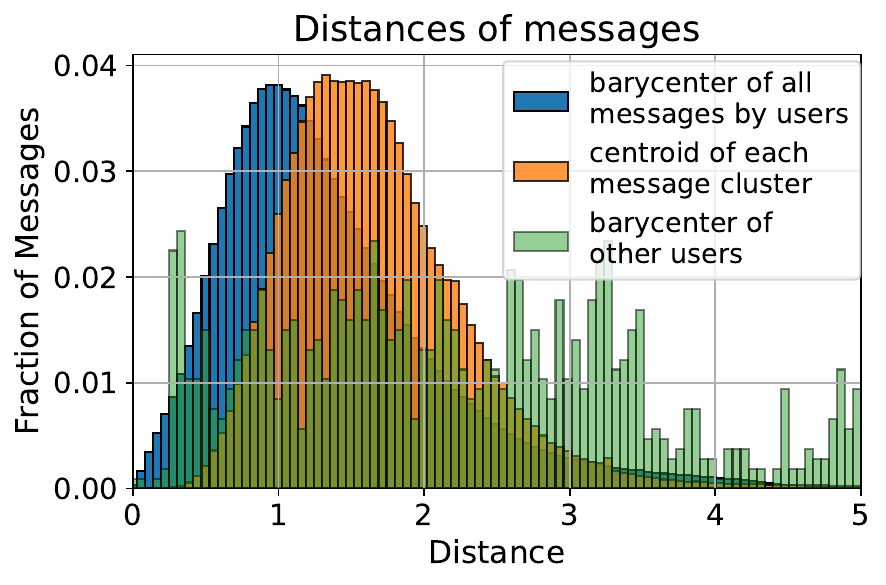}
    \caption{Distributions of Euclidean distances in the curiosity stimulus space between the messages sent by a user and (i) the barycenter of all  messages by the same user (blue), (ii) the centroid of each message's profile (orange) and (iii) the barycenter of messages sent by other users in the last 15 minutes (green).
    }
    \label{fig:distance_cluster_baricenter}
        \vspace{-0.3cm}
\end{figure}

\emph{\textbf{Takeaway}: (i) Messages are grouped into 6 clusters in the curiosity stimulus space, identifying more diverse and complex curiosity profiles, compared to  prior analysis of WhatsApp.
(ii) Clusters 0 and 1, characterized primarily by (indirect and direct) social influence, and Cluster 4, with no clear evidence of curiosity stimulation, resemble findings from the WhatsApp study,  but with a much lower presence, notably in Politics groups.
(iii) Roughly 50\% of the messages in all analyzed groups are dominated by other curiosity stimulation 
factors,  highlighting the broader and more nuanced role of curiosity stimuli in influencing user engagement.}\looseness=-1 

\subsection{User-Level Curiosity Stimulation}

After characterizing curiosity stimulation at the message level,  we shift our focus on studying it at the user level. To that end, we first analyze the distribution of messages sent by the same user within the 7-dimension curiosity stimulus space. 
In Fig.\ref{fig:distance_cluster_baricenter} we compare the distributions of the Euclidean distances of messages within this space. 
We observe that the distances between a user's messages and the user's barycenter (i.e. the average of all messages sent by the same user) in this space are statistically smaller (Mann-Whitney test, $p<0.01$) than the distances between those messages and the centroids of their respective clusters (i.e., profiles identified in the previous section). This observation suggests that the messages sent by the same user tend to be concentrated in a localized area of the stimulus space, implying that their curiosity is consistently triggered by similar types of stimuli.\looseness=-1

\begin{figure}[tttt!]
    \centering
    \includegraphics[width=
    0.8\linewidth]{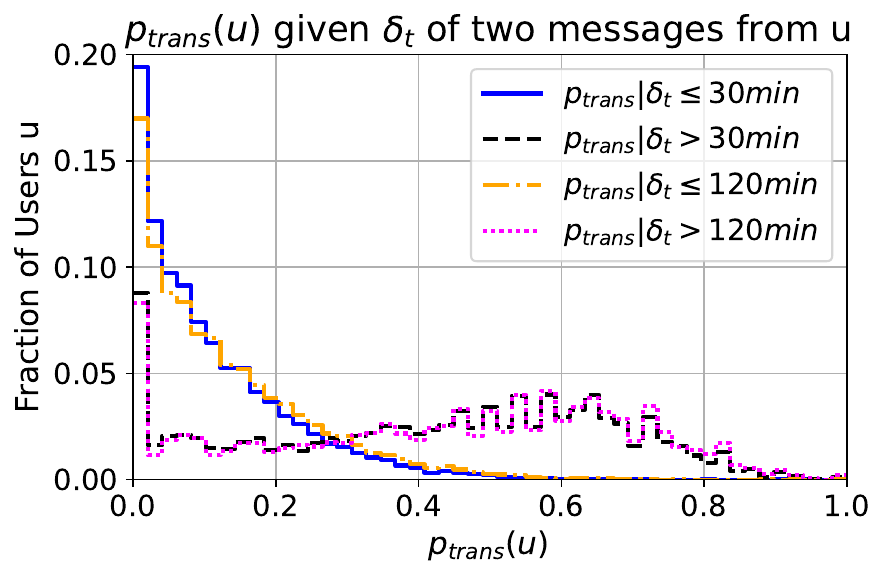}
    \caption{Distribution of the probability $p_{trans}$ that the next message from the same user in a group belongs to a different curiosity stimulus cluster, conditioned on whether the time interval $\delta_t$ between them exceeds a certain threshold.}
    \label{fig:p_trans_given_deltaT}
        \vspace{-0.3cm}
\end{figure}

We also observe that the nature of curiosity stimuli seems to be specific to each user since messages posted by different users within the same group and in close temporal proximity can exhibit markedly different stimulation profiles. This is illustrated by the green distribution in Fig. \ref{fig:distance_cluster_baricenter}, which consists of distances between a user's message and the barycenter of messages sent by other users within the preceding 15 minutes.
It is important to note that the choice of this time period is not arbitrary, but is directly tied to the selection of the window of interaction, ensuring that at least 50\% of the windows overlap. This means that users are influenced by a common context for at least half of their window of interaction.
Thus, the messages offering stimuli to them may be quite similar. Yet, the green distribution is biased towards larger distances, i.e., those users are far apart in the curiosity stimulation space, compared to the current user under analysis.\looseness=-1


\begin{figure*}[ttt!]
    \centering
    \begin{subfigure}[b]{0.25\linewidth}
        \centering
        \includegraphics[width=\linewidth]{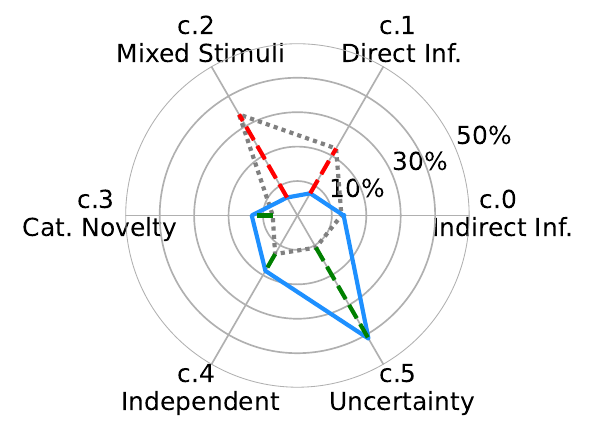}
        \caption{Politics}
        \label{fig:Politics}
    \end{subfigure}
        \hfill
    \begin{subfigure}[b]{0.25\linewidth}
        \centering
        \includegraphics[width=\linewidth]{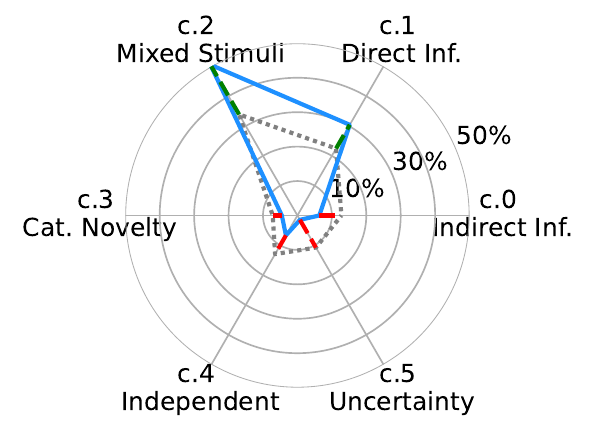}
        \caption{Cryptocurrencies}
        \label{fig:Cryptocurrencies}
    \end{subfigure}
   \hfill
    \begin{subfigure}[b]{0.25\linewidth}
        \centering
        \includegraphics[width=\linewidth]{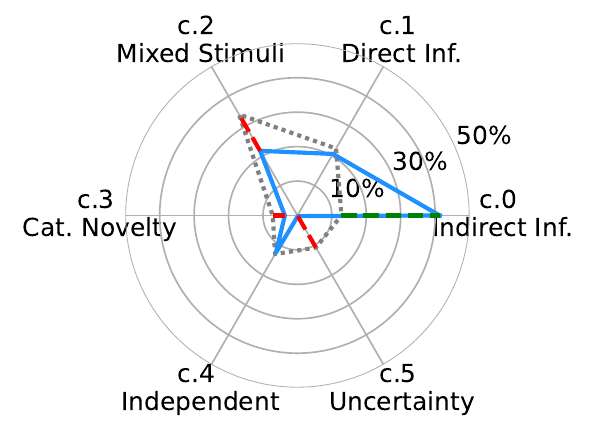}
        \caption{Linguistics}
        \label{fig:Linguistics}
            \vspace{-0.3cm}
    \end{subfigure}

    \caption{Average fraction of messages sent in each stimulus curiosity cluster from users belonging to a group of the topic (blue solid line). The dotted grey line is the average distribution over the entire population. Dashed green/red lines highlight that the distribution in the topic population is statistically higher/lower than the general one.}
    \label{fig:Radar_plot_3}
\end{figure*}

Despite such a concentration of messages from the same user in the curiosity stimulation space, we also observe some changes with time. Specifically, we measure the probability that two consecutive messages posted by the same user belong to different curiosity stimulation 
profiles, conditioned on whether the time elapsed between their postings exceeds (or not) a given threshold.  To ensure robust results, we only estimate this probability for samples (users) with at least 10 messages. 
Fig. \ref{fig:p_trans_given_deltaT} illustrates, for two such thresholds, that the probability of transitioning into a different curiosity profile increases with the time 
interval 
between them.
These results suggest that, while curiosity-driven behaviour tends to be mostly stable, it can also exhibit occasional shifts in response to
changing stimuli over time.\looseness=-1  

In general, we observe that some changes in curiosity profile are more recurrent than others. For example, when in clusters  0 and 1, for which curiosity is mostly stimulated by indirect and direct influence, respectively, users more often change to cluster 2, with mixed stimuli (roughly 40\% and 52\% of the time, respectively). For users in cluster 2, in turn, the most common transition is back to cluster 1 (35\%). 
Yet, often such transitions are only temporary fluctuations, as 23\% of transitions revert to the previous cluster in the next message, and 11\% revert after two messages. That is, despite some occasional changes, each user can be fairly well characterized by one curiosity stimulation profile that dominates his behaviour across most of the messages posted during the period of analysis\footnote{Such observation may change for longer periods of analysis. }. 
Indeed, when analyzing the messages sent by a user collectively, for 76.6\% of the users, the majority of their messages (more than 50\%) fall within a single curiosity profile.\looseness=-1

Next, we analyze deviations of user curiosity stimulation across different topics, considering the 10 group topics available in our dataset.  To that end, we take only user messages posted in groups of the same topic and compute the average (across all users)  fractions of messages in each curiosity profile. We do so for groups in each topic and compare the results against those of the overall population, so as to infer patterns that are specific to each topic. Fig. \ref{fig:Radar_plot_3} shows radar plots with the average fractions of messages (blue) in each curiosity profile (axes) for three selected topics, namely {\it Politics}, {\it Cryptocurrencies } and {\it Linguistics}. Results for the overall population are shown in grey.  Plots for the other topics are presented in Appendix A.\looseness=-1

We observe that users in groups discussing Politics (Fig. \ref{fig:Politics}) show a preference for messages driven mostly by uncertainty (cluster 5), revealing a tendency for peer discussions with a broad and balanced audience and using different media types (categories). 
This contrasts with prior work on WhatsApp, where this aspect was overlooked due to a focus solely on social influence. The same was also observed for groups in Technologies. In contrast, Cryptocurrencies (Fig. \ref{fig:Cryptocurrencies}) (as well as Bookmaking), which likely involve trust-building with individuals who have previously provided successful advice, exhibit a significantly higher proportion of messages mostly influenced by direct social influence (cluster 1). In the Linguistics groups (Fig. \ref{fig:Linguistics}), in turn, there is a marked increase in messages in cluster 0, which are strongly stimulated by indirect social influence. These results align with our manual observation that discussions in such groups are often led by language teachers who encourage conversation by posing questions and introducing subjects for debate.\looseness=-1   


Finally, we consider whether user curiosity varies based on group membership and topic. Our dataset provides limited opportunities to explore this phenomenon, as only 2\% (596) of the users are active in multiple groups, limiting the generalizability of these findings. 
Nevertheless, focusing on those users active in multiple groups, we observe that a user's dominant curiosity profile differs 52\% of the time between groups within the same topic. This percentage increases to 62\% when comparing groups in different topics. These results hint that a user's curiosity depends on the specific group they participate in, with topic diversity playing a role in shaping how their curiosity is stimulated.\looseness=-1

\emph{\textbf{Takeaway}: (i) Users' messages focus tightly around their barycenter, showing consistent curiosity-driven behavior.
(ii) Different users in the same group and timeframe may exhibit distinct curiosity profiles, emphasizing the individual nature of curiosity, even within shared environments.
(iii) While users typically follow one curiosity profile, transitions occur over time, though many are temporary.
(iv) We are the first to identify that curiosity stimulation patterns vary by topic — e.g., while uncertainty is a dominant factor driving user curiosity in \textit{Politics} groups, direct social influence is a much stronger factor in \textit{Cryptocurrencies}.}

\subsection{Active Engagement and Influence}
\label{subsec:GraphAnalysis}

\begin{figure}[tt!]
    \centering
    \includegraphics[width=0.9\linewidth]{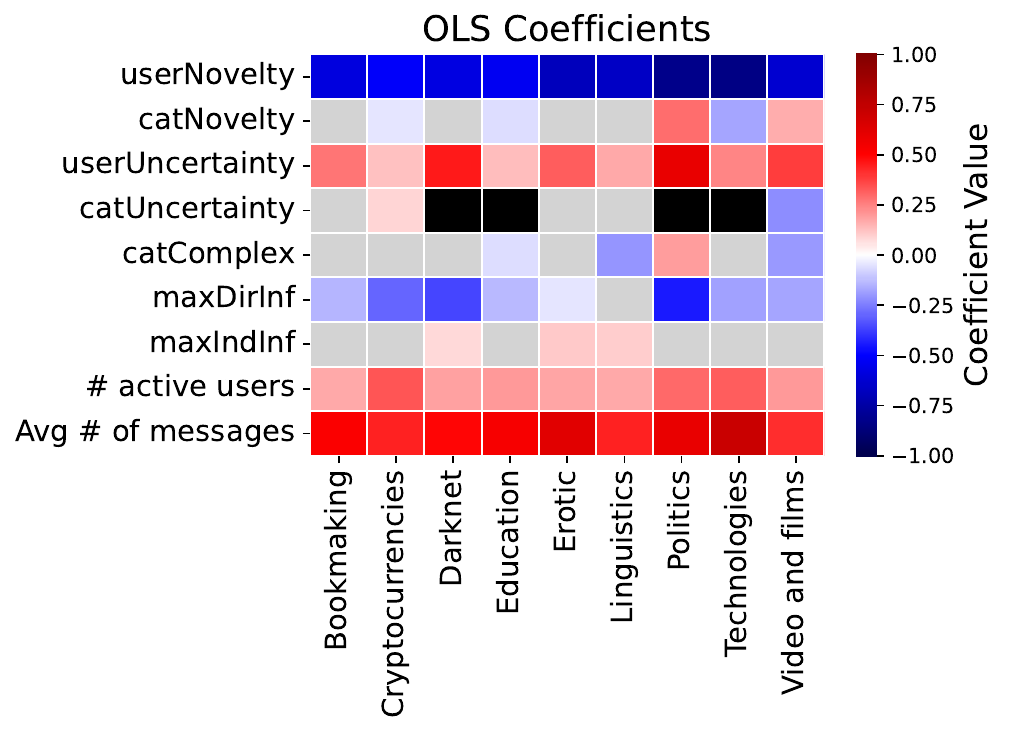}
    \caption{OLS coefficients modeling user engagement (message count) based on average curiosity metrics,  number of active users and  average number of messages by others in the group, for groups in various topics. 
   Grey cells mark non-significant coefficients ($p > 0.01$); black cells mark metrics excluded due to multicollinearity.}
    \label{fig:OLS_coefficient}
\end{figure}

To address our second research question, we study how the curiosity profile of a user relates to user engagement in group discussions. 
To the best of our knowledge, this analysis is the first of its kind in the context of instant messaging social networks, offering new insights into the role of curiosity in information spread processes.

We start by  representing the curiosity profile through the barycenter of the user's messages in the curiosity stimulus space, and we capture the user engagement by the number of messages sent by the user.
Then, we fit an OLS (Ordinary Least Squares) regression for each topic, with the number of messages sent as the response variable. As predictor variables, we use the barycenter of the user's messages in terms of curiosity metrics along with two control variables: the number of active users in the group and the average number of messages sent by other users in the group. These control variables account for the heterogeneity in group activity levels.
For each topic, we log-transform the control features and the response variable as their distributions span multiple orders of magnitude and standardize all values according to a Z-score normalization. Variables with a Variance Inflation Factor (VIF) greater than 10 are removed to avoid multicollinearity, as recommended by \citet{neter1983applied}. We opt to exclude the \textit{Software and Applications} topic due to the poor fitting (adjusted $R^2$  $<$ 0.2), focusing our analysis of the significant OLS coefficients on the remaining 9 topics.\looseness=-1 


The heatmap in Fig \ref{fig:OLS_coefficient} illustrates the OLS coefficients between the predictors (curiosity stimulation metrics and control variables) and the number of messages sent by users across different topics. Positive values  (red) suggest that higher values of certain features are associated with more frequent active engagement by the user, while negative values (blue) suggest the opposite relationship. Metrics for which the relationship is not statistically significant (grey) or that were removed due to multicollinearity  (black) are also highlighted.\looseness=-1 

We observe that both \textit{userNovelty} and \textit{maxDirInf} consistently show a negative correlation with user activity across all topics. This suggests a general pattern where users who are typically motivated by the novelty of their contributions, or by interactions with peers they frequently engage with, tend to be less actively engaged in the conversation.
In contrast, \textit{userUncertainty} exhibits a consistent positive correlation across topics, particularly in \textit{Politics} and \textit{Darknet}, indicating that users who prefer conversations with varied and balanced participation may be involved more frequently.
The other metrics show mixed effects across topics indicating the variability of certain relations by the surrounding context.\looseness=-1 

We also note the strong effect of group activity levels on user's propensity to participate.
The average number of messages sent by other users in the group shows a strong positive correlation with user engagement, suggesting that higher overall activity may encourage users to contribute more often, possibly due to the stimulating effect of a more dynamic environment. 
Similarly, the number of active users in the group also tends to have some positive influence on individual participation, across topics:  as more users actively engage in the group, individual users may perceive their contributions as part of a broader and more meaningful discussion.\looseness=-1  



\begin{figure}[t]
    \centering
    \includegraphics[width=
   0.8\linewidth]{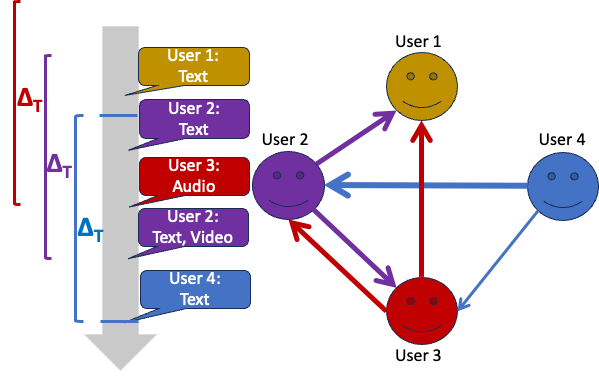}
    \caption{A toy example of an \textit{Influenced-by} network. \textit{User 4} (blue) posts a message at time t. Within the preceding time window $ [t-\Delta T,t] $, \textit{User 2} (purple) sent two messages, and \textit{User 3} (red), one. For the message posted by \textit{User 4}, we create an edge from \textit{User 4} to \textit{User 2} with  weight of $\frac{2}{3}$, and an edge from \textit{User 4} to \textit{User 3} with weight $\frac{1}{3}$. Similarly, for the second message sent by \textit{User 2}, we add two edges with weights $\frac{1}{2}$ to \textit{User 1} and \textit{User 3} (no self-loops).\looseness=-1 }
    \label{fig:toy_graph}
    \vspace{-0.3cm}
\end{figure}

As a complementary analysis of the relationship between a user's curiosity stimulation profile and the activity she engages in the groups, we study their influence on other users in the chat, in an orthogonal approach to the proposed metrics, by using graphs. More specifically, to represent the influence dynamics in group conversations, we build a {\it Influenced-by} network for each group as follows.
 We keep a  30-minute time window $\Delta T$ consistently with the window of interaction. For each message sent by user $i$ at time $t$, we draw direct edges from $i$ to all other users $j$ (different from $i$) who posted messages in the period $[t-\Delta T,t]$, that is, users  who potentially influenced $i$.  
 The weight of each edge $(i,j)$ is given by the fraction of messages sent by user $j$ out of the total messages sent during the time window. To avoid overestimating the influence of a single user who has sent multiple consecutive messages with no interleaving messages from others (\textit{chain of messages}), we count all messages in a \textit{chain} as a single one. Thus, for each message sent by $i$, the total influence from other users adds up to 1. Note that we do not add any edge if no other user sent a message within $[t-\Delta T,t]$. We illustrate the creation of an {\it Influence-by} network with a toy example in Fig. \ref{fig:toy_graph}. Finally, we merge all edges with the same source $i$ and target $j$ into one, with a weight equal to the sum of their weights.
Note that while an edge from $i$ to $j$ indicates that $i$ is influenced by $j$, it also implies that users who are more central in the network tend to exert more influence over others in the group.
We build one network for each group but focus our analyses on groups with at least 100 users in their {\it Influence-by } networks (346 groups), to ensure robust computation of graph metrics.   For each such network, we compute the PageRank centrality scores of its nodes (i.e., users).\looseness=-1 

\begin{figure}[!t]
        \centering
        \includegraphics[width=0.8\linewidth]{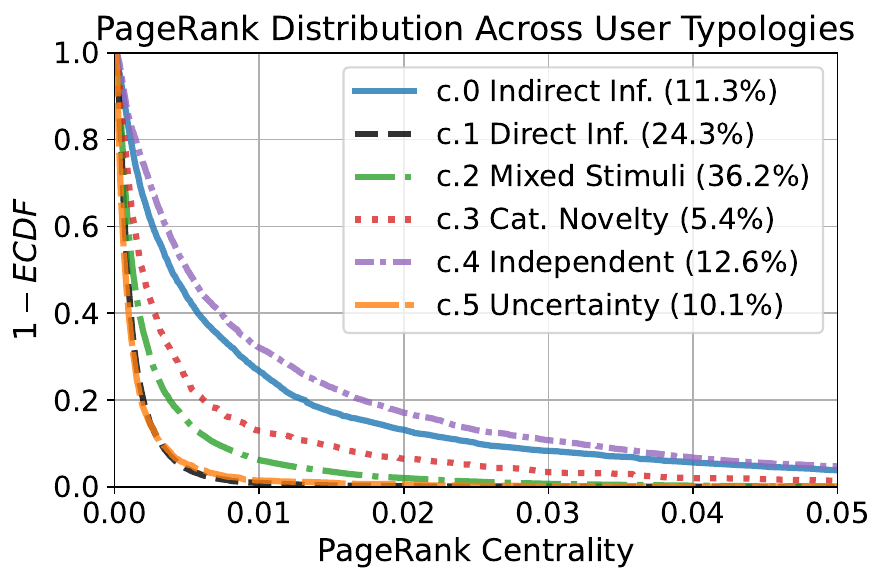}
        \caption{
       Distribution of node PageRank in the {\it Influenced-by} network, categorizing users based on the message curiosity-stimulus cluster where they hold the highest share. Percentages show how many users predominantly sent messages in each cluster. }
        \label{fig:pagerank_eccdf}

        \vspace{-0.3cm}
    \end{figure}

Recall that, aside from temporary fluctuations, the majority of a user's messages fall within the same cluster.  
This highlights the predominance of a single curiosity stimulation profile over all others (see Fig. \ref{fig:top1vstop2_rates} in Appendix A), which accounts for the 63\% of the explained variance in the distribution of the user's messages across the different curiosity stimulus profiles. 
Accordingly, we take this profile as representative of the user's curiosity stimulation and analyze the distribution of PageRank scores across users within each profile. These distributions are shown in Fig. \ref{fig:pagerank_eccdf}.\looseness=-1 

As observed, users who mainly post messages belonging to the profiles strongly driven by   \textit{direct social influence} (cluster 1) and \textit{uncertainty} (cluster 5) tend to occupy more peripheral positions within the \textit{Influenced-by} networks. This could be attributed to the fact that users with high \textit{direct social influence} tend to engage less actively, generating fewer messages  that drive ongoing discussions, thus reducing their centrality within the network. This observation is indeed consistent with the results on the OLS coefficients, shown in Fig. \ref{fig:OLS_coefficient}.
On the other hand, users characterized by high \textit{uncertainty} may see their influence diluted within the broader conversation, making them less likely to be pivotal in shaping the direction of the discourse.
In contrast, users often driven by indirect influence (cluster 0), tend to occupy more central positions in the network, suggesting that these users participate more often,  also influencing others. Indeed, as shown in Fig. \ref{fig:OLS_coefficient}, this relationship can be observed, to some degree,  for groups in some (but not all) topics, notably Darknet, Erotic and Linguistics.
Finally, it is noteworthy that users whose majority of messages are not influenced by any curiosity stimuli (i.e., independent, cluster 4)  are often the most central in the \textit{Influenced-by} networks. It is plausible to infer that such users may serve as initiators of conversations.\looseness=-1 

\emph{\textbf{Takeaway}: This study is the first to reveal key relationships between curiosity and active engagement in group discussions, notably:
(i) Users driven by novelty or recurrent peer interaction (direct social influence) tend to be less participative.
(ii) Those preferring balanced participation or indirect influence are more active and central in group interactions.
(iii) High overall group activity consistently encourages individual participation.
(iv) Independent users often occupy central roles, potentially acting as conversation initiators. \looseness=-1}

%% file: text/05_Conclusion.tex
\section{Conclusions and Future Work}
\label{sec:Conclusion}

This study investigates how the intrinsic human trait of curiosity influences user active engagement, specifically content posting, in Telegram public groups. It offers critical insights into the mechanisms that stimulate participation in group discussions. To our knowledge, this is the first research to focus on curiosity-driven engagement within Telegram.\looseness=-1

By employing different metrics which quantify aspects of curiosity stimulation, we identified 6 distinct stimulus profiles. Two profiles are primarily characterized by direct or indirect social influence, while three others are dominated by factors such as novelty, uncertainty, or a combination of multiple stimuli. By contrast, the last profile (accounting for nearly 22\% of the messages) is characterized by the absence of any of the other identified stimuli, suggesting that these messages may be driven by other factors not captured by our metrics, such as the specific content of the messages, or that they may not be influenced by curiosity at all. This underscores the complex social dynamics at play in online communities and highlights the need for further investigation into the forces driving user participation.\looseness=-1

Interestingly, we found that social influence—both direct and indirect—is a dominant component of curiosity stimulation in 28\% of all messages, particularly in groups related to topics such as {\it Bookmaking, Cryptocurrencies, Linguistics, Darknet, and Software \& Application}. 
Yet,  our results suggest that other forms of curiosity stimulation—such as novelty—also play a significant role in user participation. For example, interest in novel media types drives around 7\% of messages, while mixed stimuli account for 32\%, disputing previous studies that have only focused on social influence. 
\looseness=-1 

We also observed that as the time between messages increases, users are more likely to switch to a different curiosity stimulation profile. This supports previous findings \cite{sousa2019analyzing,Sousa_Almeida_Figueiredo_2022b} that human curiosity is dynamic and influenced by changes in both the external environment and effective stimuli. However, despite these fluctuations, the overall curiosity stimulus profile that motivates a user to post content remains relatively stable over time, considering the limited analyzed time frame. This allowed us to examine the dominant curiosity stimulation profiles at the user level across different group topics.\looseness=-1

The analysis revealed a strong relationship between specific curiosity stimulus profiles and user engagement patterns. For example, users driven by uncertainty were notably more active, particularly in groups discussing diverse and dynamic topics such as {\it Politics}, where the variety of discussions encourages more frequent participation. In contrast, users influenced by direct social social influence tended to post less often and typically occupied more peripheral roles within the group. Additionally, users who were not driven by any specific curiosity stimulus frequently emerged as conversation starters and central group figures. This finding emphasizes the importance of understanding not only individual curiosity profiles but also the broader social dynamics that influence participation and leadership within online communities.\looseness=-1

Furthermore, the diverse discussion topics present in the dataset allowed us to identify distinct patterns in curiosity stimulation based on the subject of the discussion. As noted, social influence plays a significant role in some topics, while uncertainty and other factors drive engagement in areas such as {\it Politics} and {\it Technology}. Interestingly, in groups related to {\it Video \& Films} and {\it Software \& Applications}, the independent curiosity profile dominates, suggesting that messages in these groups may not be primarily driven by curiosity.\looseness=-1

In conclusion, this research offers valuable insights into the dynamics of online communities by focusing on curiosity-driven engagement. It may enhance our understanding of the role of curiosity — a fundamental human behavioural trait linked to cognitive processes — in the context of information dissemination and spreading, aspects largely overlooked by prior works.

Future research will further explore how the content of specific messages triggers different curiosity types and how these triggers influence user engagement across various platforms. 
In this paper we prioritized a breadth-oriented approach, analyzing a diverse range of topics, groups, and languages to mitigate potential biases associated with short-term observations. We believe that examining the long-term impact of different curiosity stimuli on sustained participation and community growth will help to better understand the broader implications of these dynamics on user involvement and the evolution of digital communities, enhancing the generalizability of the findings.\looseness=-1

%% file: text/appendix.tex
\section*{Appendix A - Supplementary materials }
\label{app:A}

\begin{figure}[tbh]
    \centering
    \includegraphics[width=0.75\linewidth]{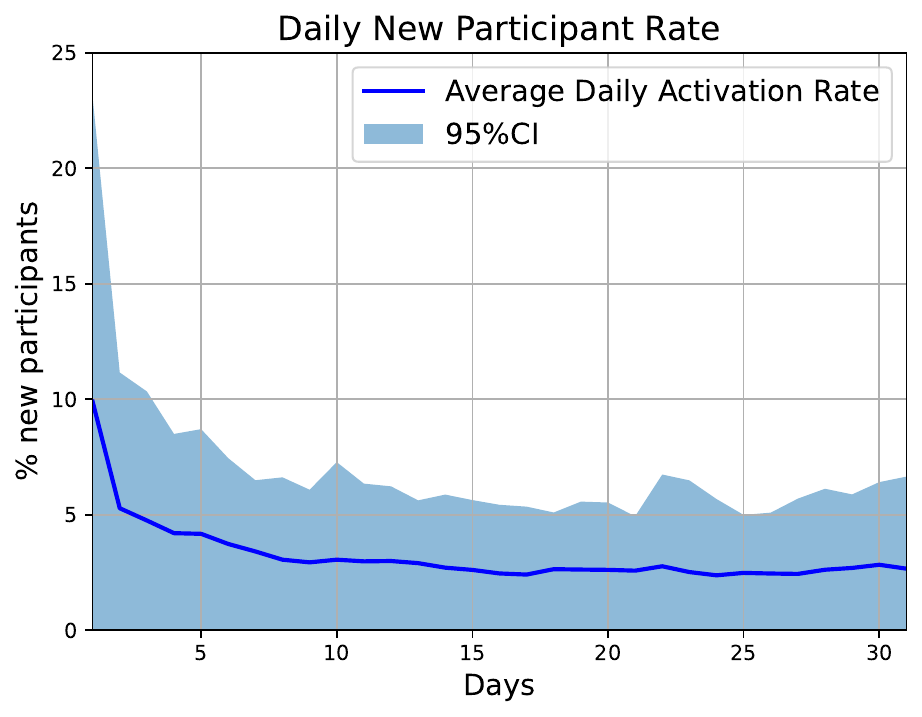}
    \caption{Percentage of users actively engaging for the first time by sending a message during the study period, shown with average values over groups and 95\% confidence intervals.}
    \label{fig:daily_activation_rate}
\end{figure}

\begin{figure}[tbh]
    \centering
    \includegraphics[width=0.75\linewidth]{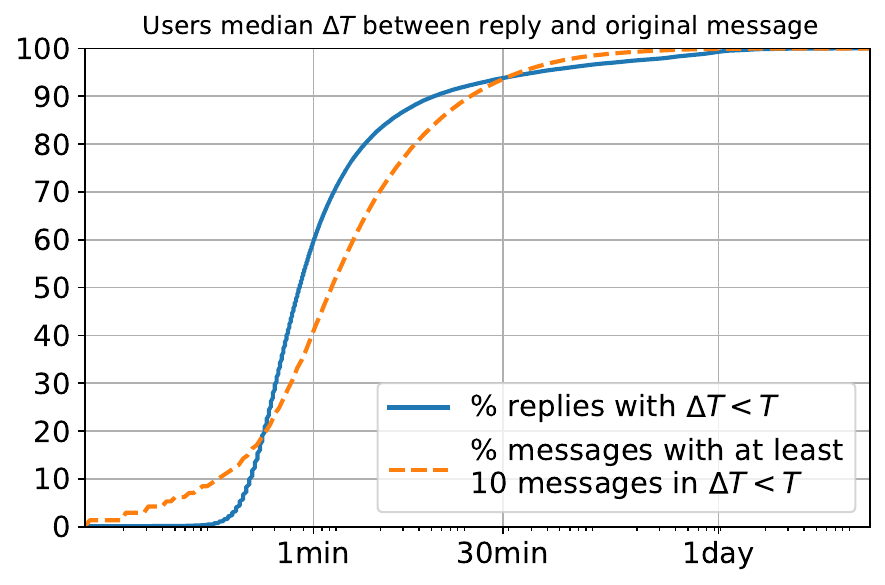}
    \caption{Cumulative distributions of user median reply times (solid blue) and the proportion of messages with at least 10 preceding messages in the same $\Delta T$ window (dashed orange). The intersection at 30 min. represents a sweet spot, balancing the trade-off between filtering noise from the user's attention period and maintaining sufficient data coverage.}
    \label{fig:reply_median_deltaT}
\end{figure}

 \begin{figure}[tbh!]
    \centering
    \includegraphics[width=
    0.9\linewidth]{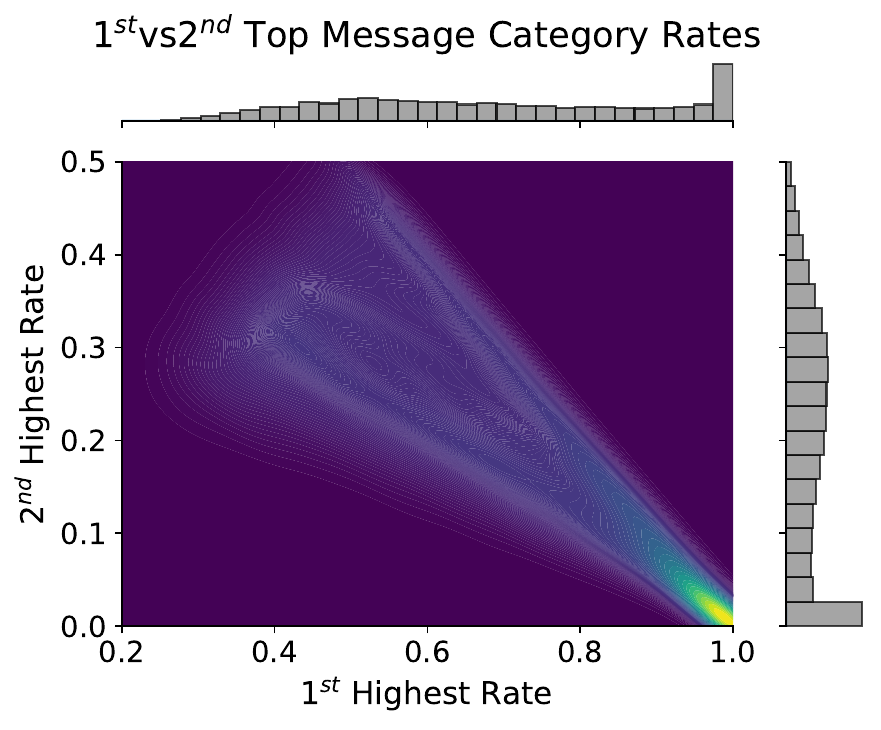}
    \caption{Share of messages in the dominant curiosity profile (where the user posted most frequently) compared to the share in the second-most active curiosity profile.}
    \label{fig:top1vstop2_rates}
\end{figure}

To complement the main findings presented in this paper, additional materials have been included in this appendix. These supplementary materials provide detailed explanations, extended analyses, and supporting results that further substantiate the claims made in the study. They aim to offer deeper insights and transparency, ensuring the reproducibility of the experiments and facilitating future research in the field.

Table \ref{tab:topic-description} offers a qualitative description of the topics, as provided by Perlo et al. (2024), and we included it here for completeness.

Figure \ref{fig:daily_activation_rate} illustrates the percentage of users actively engaging for the first time by sending a message during the study period. Although we lack information regarding the group's activity duration or the age of its members, observing the timing of users' first activation reveals an initial peak on the first day, averaging around 10\%, which then stabilizes at approximately 2.5\%.

To address the choice of the interaction time window, we analyzed user reply times and message density within varying timeframes. Using reply times as a proxy for user attention spans, we observed that over 90\% of users reply within 30 minutes. Additionally, we examined how the proportion of messages with at least 10 preceding messages changes with different time windows. The choice of a 30-minute window balances the trade-off between including too much noise (with longer windows) and filtering out excessive data (with shorter windows). This intersection of trends is illustrated in Figure \ref{fig:reply_median_deltaT}, highlighting 30 minutes as the ideal window length for our analysis.

\begin{figure*}[ht]
    \centering
    \includegraphics[width=0.65\linewidth]{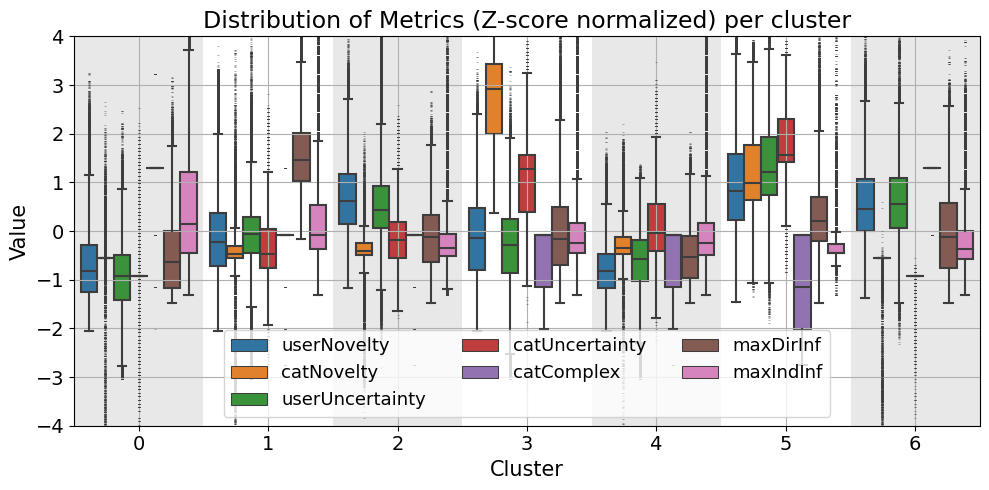}
    \caption{Boxplots of the curiosity stimulus metrics (Z-score normalized)  for message clusters.}
    \label{fig:message_cluster_boxplots}
\end{figure*}

\begin{figure*}
    \centering
    \begin{subfigure}[b]{0.45\textwidth}
        \centering
        \includegraphics[width=\textwidth]{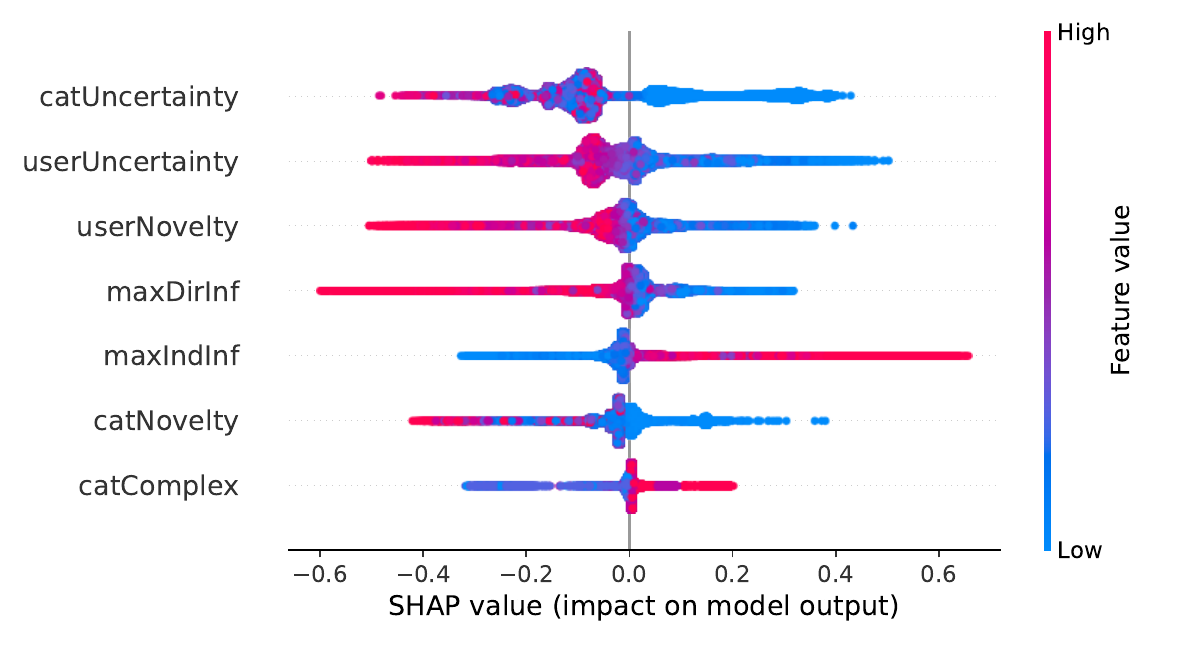}
        \caption{Cluster 0, exhibiting high indirect influence.}
        \label{fig:1}
    \end{subfigure}
   \hfill
    \begin{subfigure}[b]{0.45\textwidth}
        \centering
        \includegraphics[width=\textwidth]{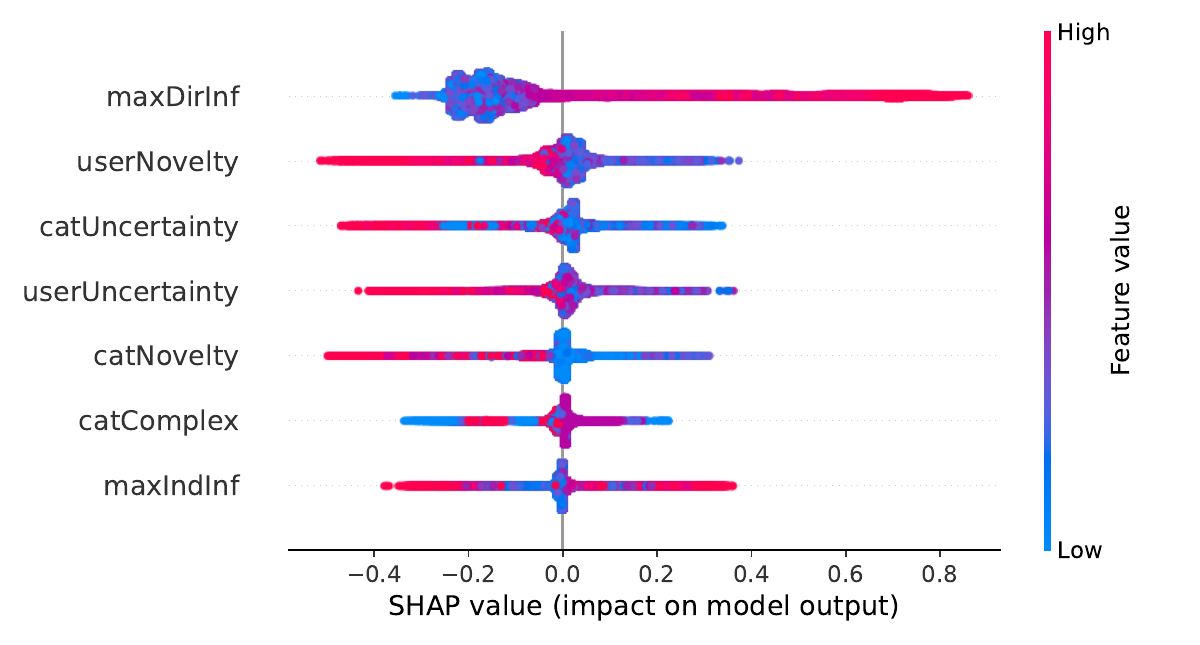}
        \caption{Cluster 1, exhibiting high direct influence.}
        \label{fig:2}
    \end{subfigure}
    
    \vskip\baselineskip
    
    \begin{subfigure}[b]{0.45\textwidth}
        \centering
        \includegraphics[width=\textwidth]{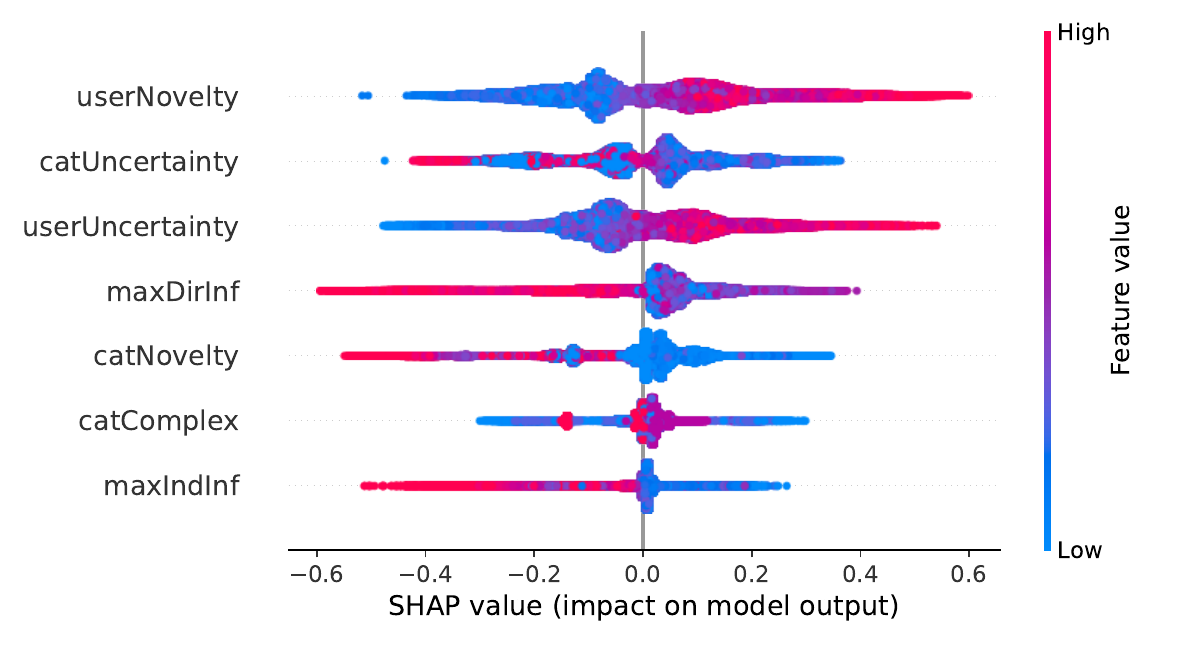}
        \caption{Cluster 2, exhibiting mixed stimuli.}
        \label{fig:3}
    \end{subfigure}
    \hfill
    \begin{subfigure}[b]{0.45\textwidth}
        \centering
        \includegraphics[width=\textwidth]{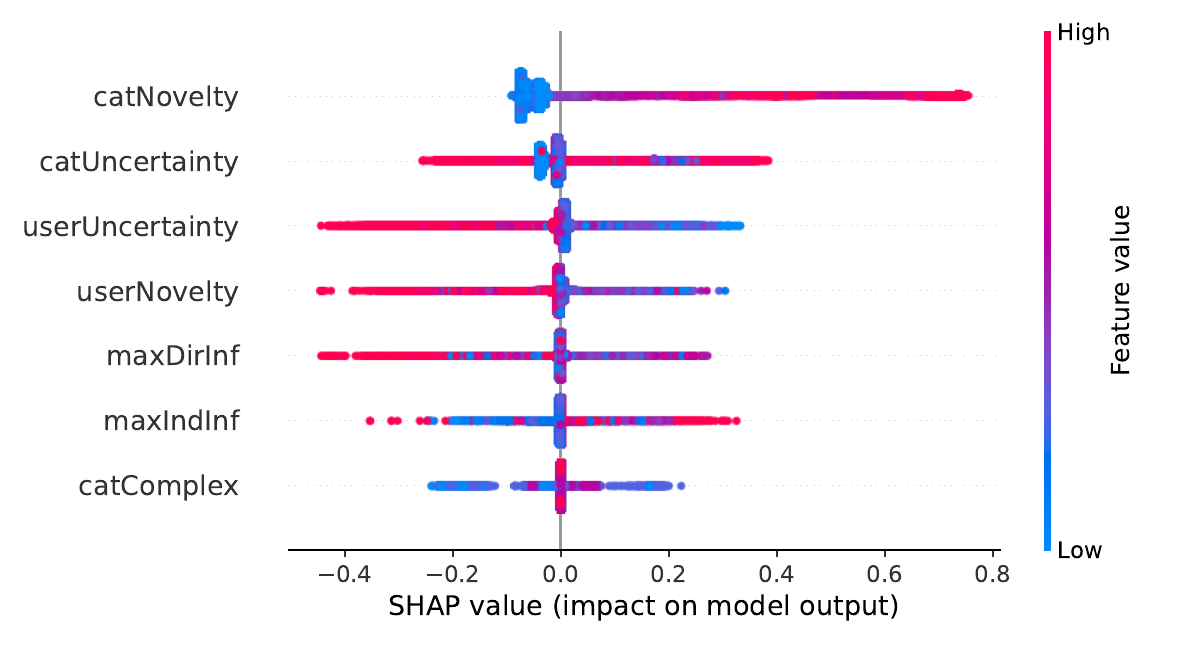}
        \caption{Cluster 3, exhibiting high categorical novelty.}
        \label{fig:4}
    \end{subfigure}
    \vskip\baselineskip
    \begin{subfigure}[b]{0.45\textwidth}
        \centering
        \includegraphics[width=\textwidth]{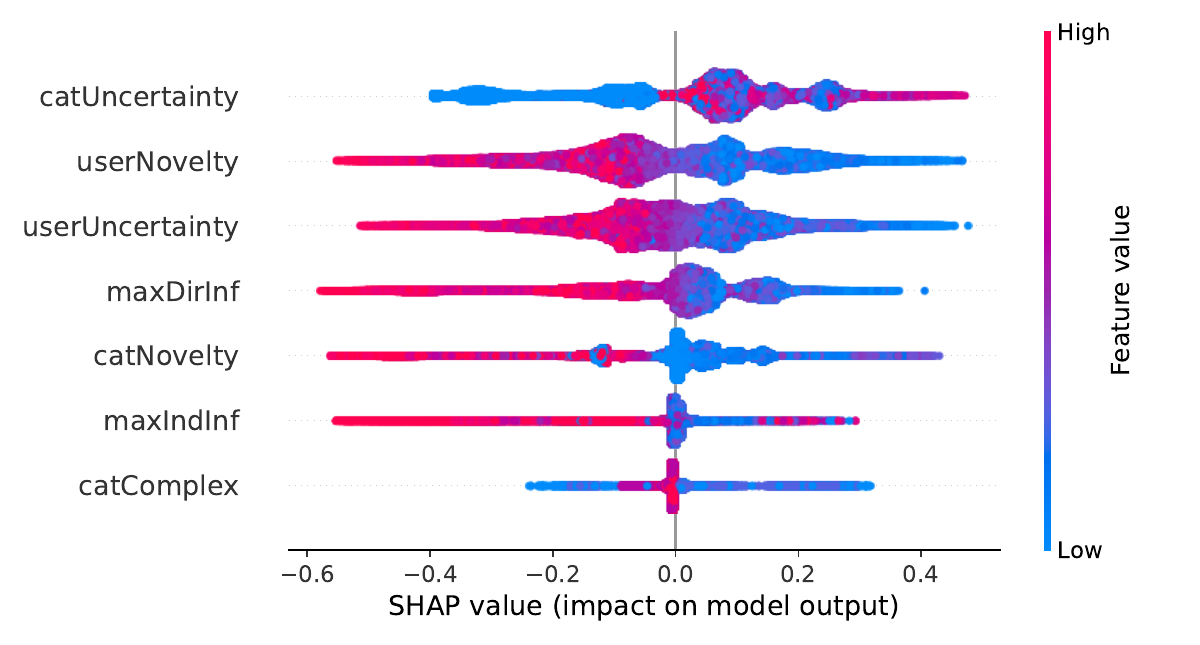}
        \caption{Cluster 4, exhibiting low stimuli values.}
        \label{fig:5}
    \end{subfigure}
    \hfill
    \begin{subfigure}[b]{0.45\textwidth}
        \centering
        \includegraphics[width=\textwidth]{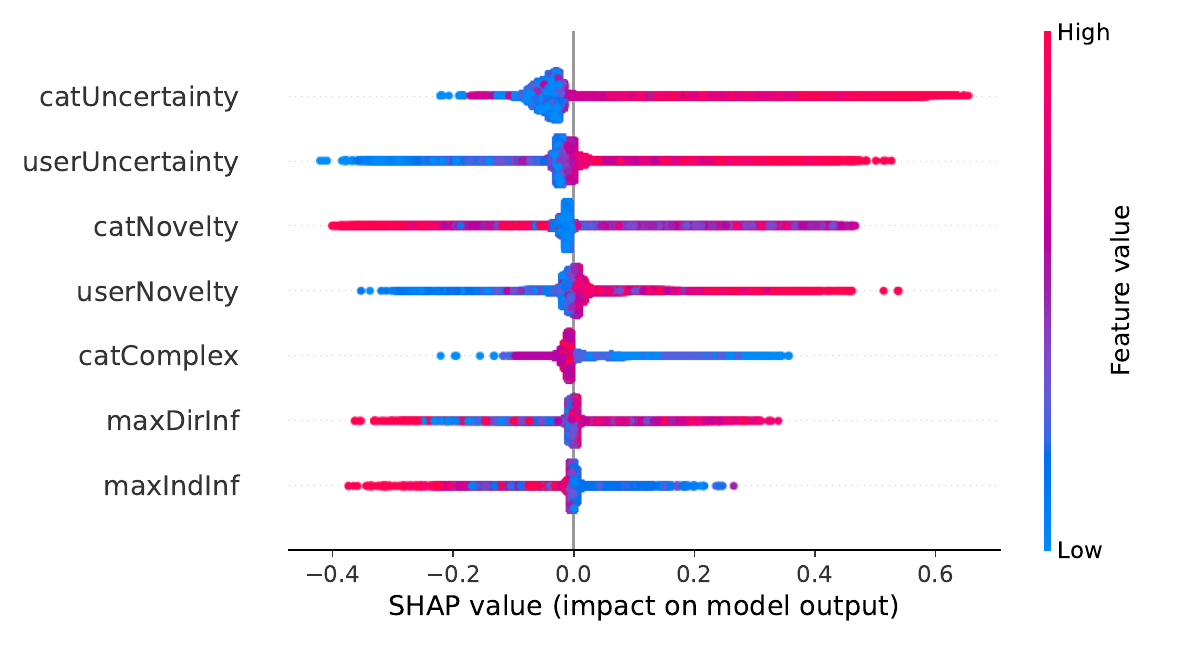}
        \caption{Cluster 5, exhibiting high uncertainty metrics.}
        \label{fig:6}
    \end{subfigure}
    
    \vskip\baselineskip
    \begin{subfigure}[b]{0.45\textwidth}
        \centering
        \includegraphics[width=\textwidth]{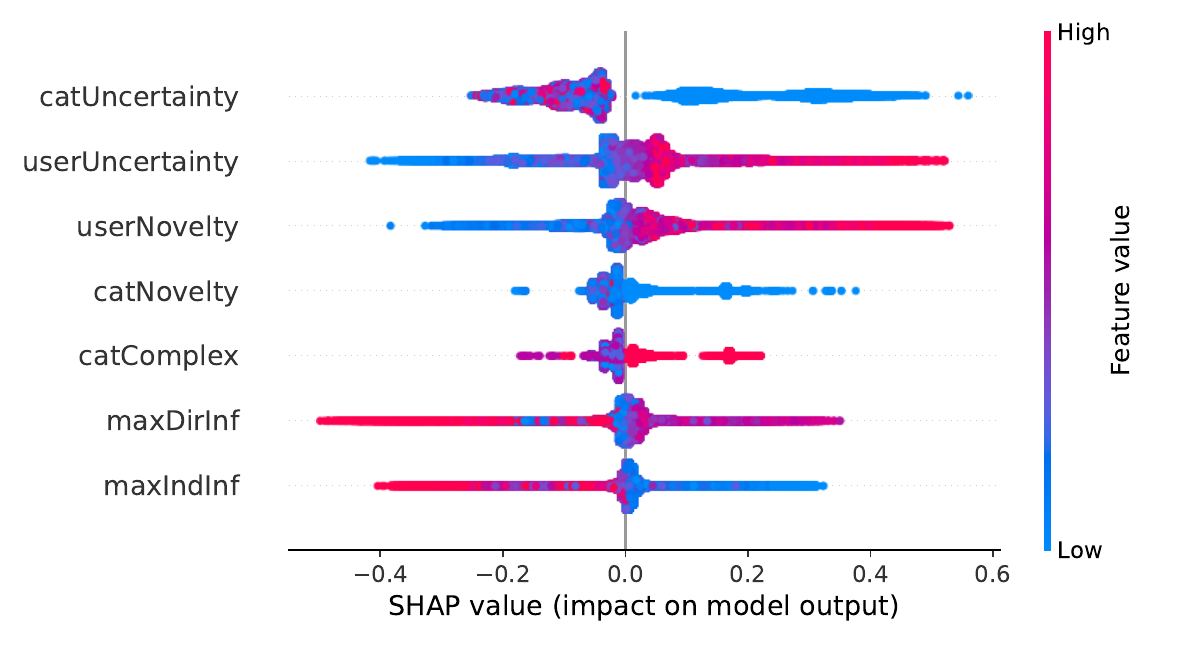}
        \caption{Cluster 6, exhibiting mixed stimuli.}
        \label{fig:7}
    \end{subfigure}

    \caption{SHAP beeswarm plot of message clusters. Features are ranked by the mean absolute SHAP values, prioritizing average impact over rare, high-magnitude effects.}
    \label{fig:main}
\end{figure*}

To enhance explainability, we present both the distribution of features for each cluster and an evaluation of feature importance. Figure \ref{fig:message_cluster_boxplots} illustrates the feature distributions (normalized Z-score) for the identified message clusters, providing insights into the distinguishing characteristics of each cluster. Additionally, we assess feature importance using SHAP values derived from a shallow decision tree. The SHAP beeswarm plot in Figure \ref{fig:main} ranks features by their mean absolute SHAP values, highlighting those with consistent contributions to the model while deprioritizing rare, high-magnitude effects.

\begin{figure*}
    \centering
    \begin{subfigure}[b]{0.33\linewidth}
        \centering
        \includegraphics[width=\linewidth]{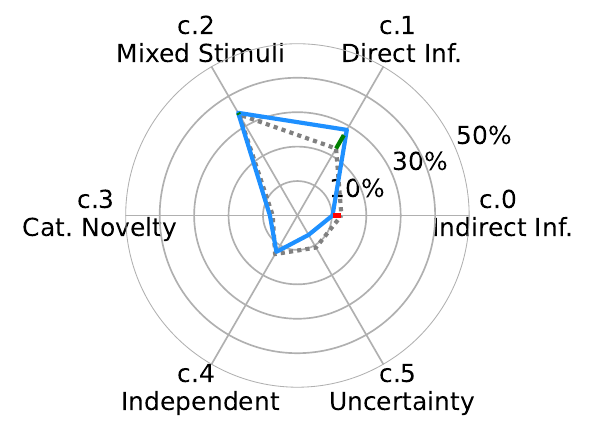}
        \caption{Bookmaking}
        \label{fig:Bookmaking}
    \end{subfigure}
\hfill
    \begin{subfigure}[b]{0.33\linewidth}
        \centering
        \includegraphics[width=\linewidth]{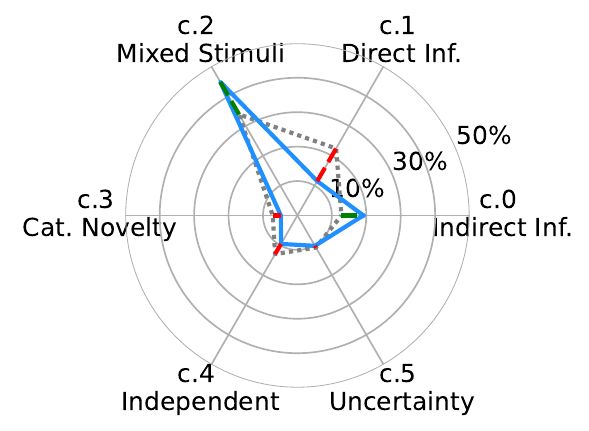}
        \caption{Darknet}
        \label{fig:Darknet}
    \end{subfigure}
\hfill
    \begin{subfigure}[b]{0.33\linewidth}
        \centering
        \includegraphics[width=\linewidth]{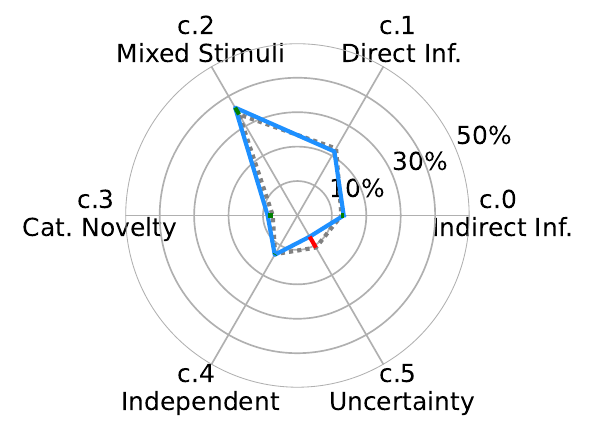}
        \caption{Education}
        \label{fig:Education}
    \end{subfigure}
\vskip\baselineskip
    \begin{subfigure}[b]{0.33\linewidth}
        \centering
        \includegraphics[width=\linewidth]{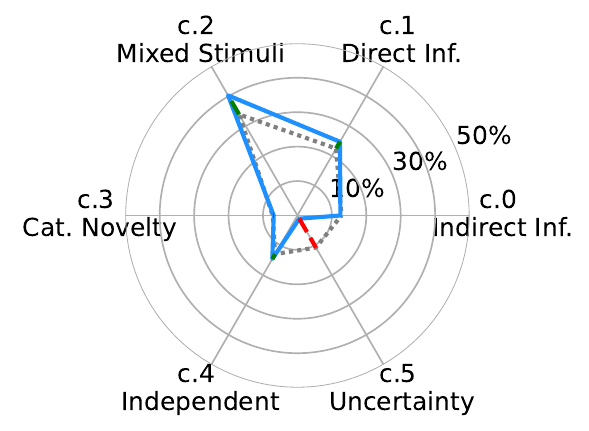}
        \caption{Erotic}
        \label{fig:Erotic}
    \end{subfigure}
\hfill
 \begin{subfigure}[b]{0.33\linewidth}
        \centering
        \includegraphics[width=\linewidth]{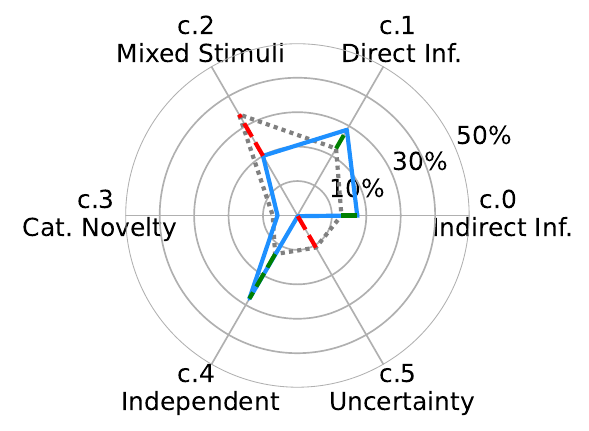}
        \caption{Software and Applications}
        \label{fig:Software & Applications}
    \end{subfigure}
   \hfill
    \begin{subfigure}[b]{0.33\linewidth}
        \centering
        \includegraphics[width=\linewidth]{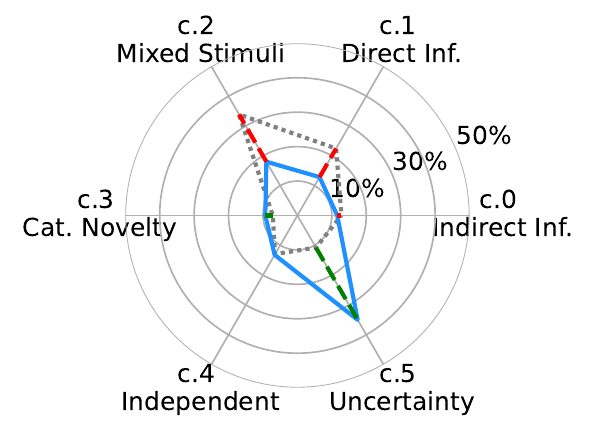}
        \caption{Technologies}
        \label{fig:Technologies}
    \end{subfigure}
    
    \vskip\baselineskip
    \begin{subfigure}[b]{0.33\linewidth}
        \centering
        \includegraphics[width=\linewidth]{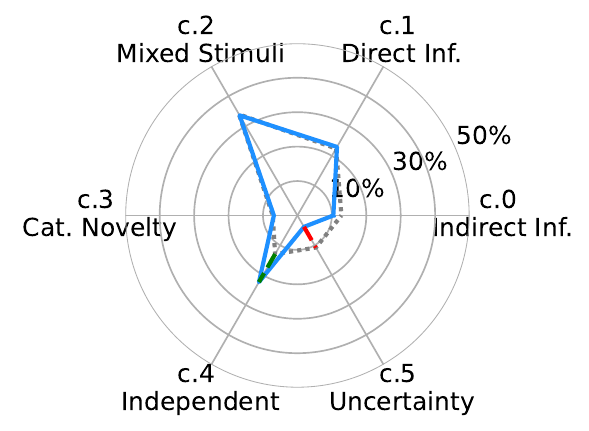}
        \caption{Video and films}
        \label{fig:Video and films}
    \end{subfigure}

    \caption{Average fraction of messages sent in each stimulus curiosity profile from users belonging to a group of the topic(blue solid line). The dotted grey line is the average distribution over the entire population. Dashed green/red lines highlight that the distribution in the topic population is statistically greater/lower than the general one.}
    \label{fig:Radar_plot}
\end{figure*}

Figure \ref{fig:Radar_plot} depicts the average fraction of messages sent by users in each curiosity stimulus profile for groups within a specific topic (solid blue line). The dotted grey line represents the average distribution across the entire population, providing a baseline for comparison. Dashed green and red lines indicate categories where the topic-specific distribution is statistically higher or lower, respectively than the general population average. This visualization highlights distinctive curiosity patterns within topic-focused groups compared to the broader population.

Figure \ref{fig:top1vstop2_rates} highlights the dominance of the most frequently used curiosity profile over the second-most active profile, and consequently over all other profiles.

\begin{table*}[htb!]
\caption{Topic description taken from \citet{perlo2024topicwiseexplorationtelegramgroupverse}.}
\centering
\begin{small}
\begin{tabular}{c p{12cm}}

\toprule
\textbf{Topic} & \textbf{Description} \\ 
\midrule
\textbf{Education} & Discussion about college and university courses and exams. \\ 
\textbf{Bookmaking} & Discussion about online betting and similar topics. \\ 
\textbf{Cryptocurrencies} & Discussion about cryptocurrencies, market stock and similar topics.  Some groups offer official support for crypto exchanges. \\ 
\textbf{Technologies} & Discussions about consumer electronics, mostly smartphones. Some groups are second-hand marketplaces for consumer electronics. \\ 
\textbf{Darknet} & Trading of mostly illegal content, i.e., credit card numbers, accounts of media platforms, etc. \\ 
\textbf{Software and apps} & Discussion about usage of software, mostly regarding Android modding and app piracy. Some groups discuss software development for specific languages or technologies.  \\ 
\textbf{Video and Films} & Discussion about movies and video sharing of movies and TV series and (possibly piracy). \\
\textbf{Politics} & Discussion about political news at large in different countries. \\
\textbf{Erotic} & Sharing and suggestion of adult content or services.  \\ 
\textbf{Linguistics} & Community of users practising a particular language for educational purposes. \\
\bottomrule
\end{tabular}
\label{tab:topic-description}
\end{small}

\end{table*}

